\documentclass[prb,twocolumn,aps,amsmath,amssymb,floatfix,showpacs]{revtex4}
\def\k{{\bf k}}
\def\q{{\bf q}}
\def\G{{\bf G}}

\def\u{{\bf u}}

\def\r{{\bf r}}
\def\d{{\bf d}}
\def\R{{\bf R}}
\def \boldtau{\pmb{\tau}}
\def\U{{\bf u}}
\def\F{{\bf F}}
\def\eps{\epsilon}
\def\vscf{{V_{\rm SCF}}}

\def\Ehxc{{E_{\rm HXC}}}

\usepackage{graphicx}
\usepackage{bm}
\usepackage{amsmath}
\usepackage{amsfonts}
\usepackage{amssymb}%
\begin{document}
\title{Adiabatic and non-adiabatic phonon dispersion in a Wannier function
approach}
\author{Matteo Calandra$^{1}$}
\author{Gianni Profeta$^{2}$}
\author{Francesco Mauri$^{1}$}
\affiliation{$^{1}$CNRS and Institut de Min\'eralogie et de Physique des Milieux condens\'es, case 115, 4 place Jussieu, 75252, Paris cedex 05, France}
\affiliation{$^{2}$Consiglio Nazionale delle Ricerche - Superconducting and Innovative Devices (CNR-SPIN),
67100 L'Aquila, Italy}

\date{\today}

\begin{abstract}
We develop a first-principles scheme to calculate adiabatic and 
non-adiabatic phonon frequencies in the full Brillouin zone.
The method relies on the variational properties of a force-constants
functional with respect to the first-order perturbation of the electronic charge 
density and  on  the localization of the deformation potential in the 
Wannier function basis. This allows for calculation
of phonon dispersion curves free 
from convergence issues related to Brillouin zone sampling.
In addition our approach justify the use of the static screened potential 
in the calculation of the phonon linewidth due to decay in electron-hole pairs.
We apply the method to the calculation of the phonon dispersion and
electron-phonon coupling in MgB$_2$ and CaC$_6$. In both compounds we demonstrate
the occurrence of several Kohn anomalies, absent in previous calculations, that are 
manifest only after careful electron and phonon momentum integration.
In MgB$_2$, the presence of Kohn anomalies 
on the E$_{2g}$ branches improves the agreement with measured phonon spectra
and affects the 
position of the main peak in the Eliashberg function.
In CaC$_6$ we show that the non-adiabatic
effects on in-plane carbon vibrations are not localized at zone center
but are sizable throughout the full Brillouin zone. 
Our method opens new perspectives in large-scale first-principles 
calculations of dynamical properties and electron-phonon interaction. 
\end{abstract}
\pacs{ 63.20.dk, 71.15.-m, 63.20.kd }

\maketitle

\section{Introduction}

Electron-phonon (EP) interaction is responsible of many important phenomena in
solids. As an example, the temperature behavior of the electron relaxation time in
metals is to a great extent due to the scattering between carriers and atomic
vibrations \cite{Ziman} such that finite temperature transport is largely ruled by the EP
interaction. Similarly, the high temperature heat capacity in metals is enhanced by
the increased electronic mass due to the interaction with lattice vibrations 
\cite{Grimvall}. In
metals, at low temperatures, EP coupling can generate a superconducting state in
which electrons move with no electrical resistance \cite{Schrieffer}. 
It also can increase the effective
mass of the carriers so much that the system is driven from a metallic to an insulating
state, as it happens in case of polaronic or Peierls instabilities \cite{Peierls}. 
Finally, the electron-phonon scattering is often the largest source of phonon 
damping in phonon-mediated superconductors \cite{Shukla}.

First-principles theoretical determination of the electron-phonon coupling strength in solids requires
the calculation of the electronic structure, the vibrational properties, 
and the electron-phonon coupling matrix elements. 
In state-of-the-art electronic structure calculations
these quantities are obtained using the adiabatic Born-Oppenheimer approximation
\cite{Born27} and density-functional theory (DFT) in the linear-response approach 
\cite{pwscfRMP,Savrasov,GonzePhonon}.

More specifically, within the Born-Oppenheimer approximation, 
the determination of phonon frequencies, that
are related to the real part of the phonon self-energy,
requires the calculation, {\it in a self-consistent manner}, of the
variation of the Kohn-Sham potential $\vscf$ with respect to
a {\it static} ionic displacement $\U$ \cite{pwscfRMP},
namely ${\delta \vscf(\r)\over \delta \U}$.
As the displacement of the ions is static, the obtained
${\delta \vscf(\r)\over \delta \U}$ is real. 
Conversely, the phonon linewidth, related to the imaginary part of the phonon
self-energy, is obtained in a {\it non self-consistent} procedure
using the electron charge density and the previously determined 
${\delta \vscf(\r)\over \delta \U}$.
The advantage of using a non self-consistent procedure to study
phonon linewidth is mainly related to the less expensive computational load  
with respect to a self-consistent one.
In addition, as recently demonstrated,  interpolation schemes
\cite{GiustinoPRB,Eiguren} of the electron-phonon matrix elements 
can be used to calculate the imaginary
part of the phonon self-energy on ultra-dense k-point grid.

It is however unclear to what extent this procedure of calculating
self-consistently the real part of the phonon self-energy and non
self-consistently the imaginary part is actually correct.
Indeed, the proper way of treating phonons should be to consider a 
monochromatic time-dependent displacement (at the phonon frequency $\omega$) 
and perform a time-dependent self-consistent linear-response scheme.
The resulting variation of the self-consistent potential 
${\delta \vscf(\r,\omega)\over \delta \U}$ would then be a complex
quantity.
The real part of the resulting phonon self-energy would then determine the phonon frequencies while its
imaginary part would lead to the phonon linewidth. 
In this way, non-adiabatic (in the sense of Ref. \onlinecite{Saitta}) {\it dynamical} phonon frequencies could be accessed.
Although feasible in principle this procedure  would require a
full rewriting of the linear response code 
including time dependence and it would also be more
expensive then a standard static linear-response calculation.
Moreover, in the presence of Kohn anomalies and long-range force-constants,
where extremely accurate  k-point sampling of the Fermi surface is needed to converge
the phonon self-energy, the calculation would  be unfeasible.

Thus, it would be desirable to have a non -self consistent linear-response formulation to obtain both the real and imaginary phonon self-energy, both within the adiabatic/static and non-adiabatic/dynamic  approximation. 

In this work we develop a scheme to calculate non self-consistently
both the real and the imaginary part of the phonon self-energy
by using a functional that is variational with 
respect to the variation of the self-consistent charge density. 
Our method opens the way to calculate adiabatic/static and non-adiabatic/dynamic phonon frequencies 
using an ultra-dense sampling of the electron and phonon wave-vector in a non 
self-consistent way, starting from the static self-consistent variation 
of the time independent Kohn-Sham potential (${\delta \vscf(\r)\over \delta \U}$), 
obtained with a coarse sampling. 
In this way, the main computational load related to 
phonon frequencies calculation is drastically reduced.
The efficiency of the method is further enhanced by using the interpolation of the 
electron-phonon matrix elements \cite{GiustinoPRL} based on Wannier 
functions \cite{Marzari}. 

The method is applied to study the dynamical and superconducting properties of  MgB$_2$ and CaC$_6$, two of the most studied
superconducting materials in the last few years.
These applications are meaningful and computationally challenging. In fact, although many experimental and theoretical studies appeared, there are still important open issues and debated results. 

For example, conflicting results for the calculate MgB$_2$  electron-phonon coupling ($\lambda$)  
are present in the literature mainly  due to difficulties in Brillouin Zone sampling. 
This, apart its fundamental importance,  prevents a full understanding of the normal and superconducting properties of this material.

In the case of CaC$_6$, very large non-adiabatic effects were predicted\cite{Saitta} and measured\cite{DeanNA} 
at zone-center. It is unclear to what extend non-adiabatic effects are sizable far from $\Gamma$-point.
In this last case, the adiabatic effects can be relevant for thermodynamic properties
resulting from an average of the phonon frequencies over the Brillouin zone.
 
The paper is organized as follows. 
In Section \ref{section:theory} we derive the variational formulation of the 
force-constant matrix.
This will be done starting from the formal definition of force constants in the case 
of a monochromatic time-dependent ionic displacement (Sec.\ref{sec:tha}).
In Sec. \ref{sec:adiabaticnonadiabatic} we discuss when non-adiabatic/dynamic effects
should be expected in the phonon spectra.
The linear response equations for the dynamical matrix are
 introduced (Sec.\ref{sec:thb}) in the
general case and then 
specialized in the density functional formulation (Sec.\ref{sec:thc}).

In Section \ref{sec:varph}, we develop the variational formulation of 
the force constants in the density functional linear-response theory and outline 
the computational framework of phonon and electron-phonon coupling 
(Sec.\ref{sec:appr_fun} and \ref{sec:pract1}).
In Sec. \ref{sec:interp_lambda} we show how our approach 
justify the use of the static screened potential 
in the calculation of the phonon linewidth due to decay in electron-hole pairs.
Section \ref{sec:wannier} will be devoted to the description of the implementation of the theory using the Wannier interpolation scheme.

In Section \ref{sec:appli} we reports results on dynamical and superconducting properties of MgB$_2$ (Sec.\ref{mgb2}) and CaC$_6$ (Sec.\ref{cac6}).

Section \ref{conclusions} summarizes our conclusions.

\section{Theory}
\label{section:theory}

\subsection{Time dependent dynamical matrix and phonon damping.\label{sec:tha}}

We consider a crystal with N atoms in the unit cell submitted to a time dependent perturbation, in which the position of an atom is identified by the vector 
\begin{equation}
\R_I \equiv \R_L + \boldtau_{s} + \U_{I}(t), 
\end{equation} 
where  $\R_L$
is the position of the $L$-th unit cell in the Bravais lattice,
$\boldtau_s$ is the equilibrium position of the $s$-th atom in the
unit cell,  $\U_{I}(t)$ indicates the deviation from equilibrium of
the nuclear position and $I=\{L,s\}$. 
The force at time $t$ acting on the $J$-th nucleus ($J=\{M,r\}$) due to the displacement $\U_{I}(t')$ of the atom $I$-th at time $t'$ is 
labeled $\F_{J}(t)$.
The force constants matrix is defined as:
\begin{equation}
C_{IJ}(\R_L-\R_M;t-t')=-\frac{\delta \F_{J}(t)}{\delta \U_{I}(t')}
\end{equation}
where we used the translational invariance of the crystal and make evident the dependence of $C_{IJ}$ on the lattice vector 
$\R_L-\R_M$ 
(to lighten the notation we omit it in the following equations where no confusion may arise). 
The $\omega$-transform of the force-constants matrix is thus:
\begin{equation}
C_{IJ}(\omega)=\int dt e^{i\omega t}C_{IJ}(t)
\end{equation}
While the force-constants matrix $C_{IJ}(t)$ is a real quantity,
its $\omega$-transform $C_{IJ}(\omega)$ is not real and has both a real
and imaginary part.
The Fourier transform of the force-constant matrix is
\begin{equation}
C_{sr}(\q,\omega)=\sum_{L} e^{-i\q\R_L} C_{Ls,Mr}(\omega)\label{eq:cft}
\end{equation}
where, without loss of generality, we have chosen $\R_{M}={\bf 0}$.
The Hermitian and anti-Hermitian combination of the force-constant matrix
in momentum space are:
\begin{eqnarray}
D_{sr}(\q,\omega) =\frac{1}{2\sqrt{M_s M_r}}
\left[C_{sr}(\q,\omega)+C_{rs}(\q,\omega)^{*}\right]\label{eq:redynmat}\\
A_{sr}(\q,\omega) =\frac{1}{2 i \sqrt{M_s M_r}}
\left[C_{sr}(\q,\omega)-C_{rs}(\q,\omega)^{*}\right]\label{eq:imdynmat}
\end{eqnarray}
where $M_s$ is the mass of the s-th atom in the unit cell. 
These quantities are associated to the real and imaginary part of the 
dynamical matrix in coordinate space.
If the imaginary part of the dynamical matrix 
is small with respect to its real part, namely
\begin{equation}
|A_{sr}(\q,\omega)| << |D_{sr}(\q,\omega)|
\label{eq:im_neg}
\end{equation}
then the self-consistent condition
\begin{equation}
\det\left|D_{sr}(\q,\omega_{\q\nu})-\omega_{\q\nu}^2\right|=0
\label{eq:diagdyn}
\end{equation}
determines non-adiabatic/dynamic phonon frequencies $\omega_{\q\nu}$ and
phonon eigenvectors $\left\{{\bf e}^{s}_{\q\nu}\right\}_{s=1,N}$ 
and $\nu=1, 3N$ indicates the phonon branches.
The adiabatic/static phonon frequencies and eigenvectors are obtained considering a 
static perturbation, 
thus diagonalizing $D_{rs}(\q,\omega_{\q\nu}=0)$.

On the other hand, the imaginary part of the force-constants matrix determines the phonon-damping 
\begin{equation}
\gamma_{\q\nu}=\frac{2}{\omega_{\q\nu}}
\sum_{s,r}{\bf e}_{\q\nu}^{s}A_{sr}(\q,\omega_{\q\nu}){\bf e}_{\q\nu}^{r}
\label{eq:lw_C}
\end{equation}
and  the phonon linewidth.

\subsection{Adiabatic and non-adiabatic phonons\label{sec:adiabaticnonadiabatic}}

In the previous section we have defined the non-adiabatic/dynamic phonon frequencies as the eigenvalues of the Hermitian part of the time dependent 
dynamical matrix, and the adiabatic/static phonon frequencies as the eigenvalues of the static time-independent dynamical matrix. 
Solid-state text-books\cite{Ashcroft,Ziman} and first-principles calculations of the phonon dispersion\cite{pwscfRMP,deGironcoli95,GonzePhonon,Quongphonon,Savrasovphonon}, usually treat only the adiabatic/static case, since
it is commonly assumed that the adiabatic phonon frequencies coincide with the non-adiabatic ones. 

In insulators, where the fundamental energy gap between the electronic ground state and the first available excited state is much larger than the phonon energy, the adiabatic/static approximation is well justified. In metals the situation is more complex.\cite{Maksimov,Saitta,CalandraPhysC}

The crucial parameter in metals is the electron relaxation time $\tau$.
In absence of electron-defect, electron-electron and electron-phonon scattering, $\tau$ is infinite.
In a real metallic system, the presence of these scattering processes results
in a finite relaxation time $\tau$.  We can 
define three cases:
(i) a clean limit, when the electron relaxation time $\tau$ is much larger than the phonon period divided by $2\pi$
(ii) a dirty limit, when the electron relaxation time $\tau$ is much smaller than the phonon period divided by $2\pi$ 
(iii) an intermediate regime, when the electron relaxation time $\tau$ is comparable to the  phonon period divided by $2\pi$.

It has been shown\cite{Maksimov,Saitta,CalandraPhysC} that in the dirty limit 
the non-adiabatic phonon frequencies coincide with the adiabatic ones also 
in metals. Instead, in the clean limit and in the intermediate regime, the adiabatic and non-adiabatic frequencies are, in general, different.

The non-adiabatic calculations based on time-dependent DFT (described in the following sections) are performed in the perfect clean limit. Indeed, in our time-dependent 
calculations, the electron relaxation time $\tau$ is infinite, since 
 we use an instantaneous (real in the frequency space) exchange-correlation Kernel, and we do not 
consider a broadening of the electronic levels due to defects and electron-phonon scattering.

Thus, our non-adiabatic/dynamic DFT frequencies should be used to reproduce the phonon frequencies measured in metal in the clean-limit. 
Instead, the adiabatic/static DFT frequencies (those generally computed with DFT linear response codes) should be used to reproduce the phonon frequencies measured in a metal in the dirty limit.  
Finally, to reproduce the phonon frequencies measured in a metal in the intermediate regime, one should explicitly include electron-lifetime effects in the linear response calculation, as it has been done for zone center phonon in Refs. ~\onlinecite{Saitta,Cappelluti}.

By considering a single band and by linearizing the electronic-band dispersion near the Fermi energy, it has been shown\cite{Maksimov} that the differences between the adiabatic and non-adiabatic phonons are largest at the center of the Brillouin zone (BZ) and vanish for $q\gg \omega_{\q\nu}/{v}_{\rm F}$, where ${v}_{\rm F}$ is the Fermi velocity. Such differences, at the BZ center and in the clean limit,
have been computed within DFT\cite{Saitta}, and can be very sizable (up to 30\% of the phonon frequency).  However a detailed DFT study of non-adiabatic effect away from the BZ center (beyond a linearized one band approximation\cite{Maksimov}) is still missing.

\subsection{Time dependent linear response theory\label{sec:thb}}

The force-constant matrix can be evaluated in the linear response theory, considering that 
the  atomic displacement induces a perturbation in the external potential acting on the electrons.

The Hellmann-Feynmann theorem \cite{Hellmann37,Feynman39, DiVentra} states that  the force on atom $J$, $\F_J(t)$ can be evaluated 
in terms  of the variation of the external potential:
\begin{eqnarray} 
\F_J(t)
&=& -\int d\r \,n(\r,t)
{\delta V_\mathrm{ext}(\r) \over \delta \R_J}
\label{eq:F_t}
\end{eqnarray}
where $n(\r,t)$ is the electronic charge density and
$V_\mathrm{ext}(\r)$ is the external potential, namely:
\begin{equation}
V_\mathrm{ext}(\r)=-\sum_I \frac{Z}{|\r-\R_I|}+\frac{1}{2}\sum_{I\ne J}
\frac{Z_I Z_J}{|\R_J-\R_I|}
\label{eq:Vext}
\end{equation}
It is worthwhile to recall that  $V_\mathrm{ext}(\r)$ does not depend
explicitly on time but only through the dependence on time of the
phonon displacement $\U_{I}(t)$. 
In linear response the force-constant matrix $C_{IJ}(\omega) $
is written as\cite{pwscfRMP}:
\begin{eqnarray}
C_{IJ}(\omega)={  \int {\delta n(\r,\omega) \over \delta \U_{I}}}
{\delta V_\mathrm{ext}(\r) \over \delta \U_{J}}  d\r \nonumber \\
+ \int n_0(\r)  {\delta^2 V_\mathrm{ext}(\r) \over \delta \U_{J}\delta \U_{I}} d\r . 
\label{eq:hessian} 
\end{eqnarray}
where $n_0(\r)$ is the unperturbed charge density, $ n(\r,\omega)$ is the $\omega$-transform of the time-dependent charge density 
$n(\r, t)$ and 
${\delta V_\mathrm{ext}(\r) \over \delta \U_{J}} $ and ${\delta^2 V_\mathrm{ext}(\r) \over \delta \U_{I}\delta \U_{J}} $
are real and time-independent quantities evaluated at the equilibrium 
position of the nuclei. 
On the contrary 
${\delta n(\r,\omega) \over \delta \U_{I}}$ is complex.

\subsection{Time-dependent linear response in density functional theory\label{sec:thc} }

In this section, we examine the linear-response calculation of the real-space force-constants
in the framework of density functional theory. The derivations are kept in real space
as we want to keep the presence of imaginary terms (damping) in the force-constants matrix as manifest.

In order to compute $n(\r,\omega) $,  we assume a monochromatic perturbation of the form
\begin{equation}
\U_{I}(t)=\U_{I}(\omega) (e^{i\omega t}+e^{-i\omega t})
\end{equation}
where $\U_{I}(\omega)$ is real.
In density functional theory the resulting monochromatic perturbing potential 
is the external potential
$V_\mathrm{ext}(\r)$ in Eq. \ref{eq:Vext} . 
The derivative of the $\omega$-transform of the charge density
with respect to a ionic displacement
is written as\cite{pwscfRMP}
\begin{eqnarray}
&&n_{I}^{1}(\r,\omega,T)={\delta n(\r,\omega)\over \delta \U_{I}} = 
2 \sum_{\k i, \k^{\prime} j}^{N_{k}(T)} (f_{\k i}(T) -f_{\k^{\prime} j}(T))\nonumber \\
& &\times\frac{\langle \psi_{\k^{\prime} j}|\delta
  \vscf(\r,\omega)/\delta \U_{I}|\psi_{\k i}\rangle }{\eps_{\k
    i}-\eps_{\k^{\prime} j}+\omega+i\eta}
\psi_{\k i}^{*}(\r) \psi_{\k^{\prime} j}(\r)
\label{eq:dnomega} 
\end{eqnarray}
where $\k$,$\k^{\prime}$ label the crystal momentum, $i$,$j$ are
band indexes, the $T$ is the electronic temperature (or
broadening) in the Fermi function $f_{\k i}(T) $, $\eta$ is
an arbitrarily small positive real number, $N_k(T)$
is the number of k-points needed to converge the sum in Eq. \ref{eq:dnomega}
at the electronic temperature $T$ and the factor 2 (here and in the following) accounts for the spin-degeneracy.
Finally, $V_\mathrm{SCF}$ is the Kohn-Sham self-consistent potential.
The quantity $\delta \vscf(\r,\omega)/\delta \U_{I}$ is complex and
given by
\begin{eqnarray}
{\delta \vscf(\r,\omega) \over \delta \U_{I}}=
{\delta V_\mathrm{ext}(\r) \over \delta \U_{I}}+
\int K(\r,\r') n_{I}^{1}(\r,\omega,T)d\r^{\prime}.\nonumber \\
\label{eq:Dvscf_DR}
\end{eqnarray}
where $K(\r,\r')={\delta \Ehxc[n]\over \delta n(\r) \delta n(\r^{\prime})}$
is the kernel of the Hartree and Exchange and correlation functional,
$\Ehxc[n]$.
As usual we have assumed the Hartree and exchange-correlation Kernel 
to be instantaneous (real in $\omega$ space).

Substitution of Eq. \ref{eq:dnomega} in Eq. \ref{eq:hessian} leads to:
\begin{eqnarray}
& &C_{IJ}(\omega,T)=2 \sum_{\k i, \k^{\prime} j}^{N_{k}(T)} (f_{\k i}(T) -f_{\k^{\prime} j}(T))
\nonumber \\
&\times&\frac{\langle \psi_{\k^{\prime} j}|\delta
  \vscf(\r,\omega)/\delta \U_{I}|\psi_{\k i}\rangle 
\langle\psi_{\k i}|\delta V_\mathrm{ext}(\r) / \delta \U_{J} 
|\psi_{\k^{\prime} j}\rangle}{\eps_{\k
    i}-\eps_{\k^{\prime} j}+\omega+i\eta}
   \nonumber \\
& &\,\,\,\,\,\,\,\,\,\,+ \int n_0(\r)  {\delta^2 V_\mathrm{ext}(\r) \over \delta \U_{J}\delta \U_{I}} d\r . 
\label{eq:Cvscfvb} 
\end{eqnarray}
where, from now on, we explicitly indicate the dependence on the electronic temperature
$T$.
This expression of the force-constants matrix is normally used in 
standard implementations of linear-response theory \cite{pwscfRMP}.

Further substitution of Eq. \ref{eq:Dvscf_DR} in 
Eq. \ref{eq:Cvscfvb} gives the following alternative, but equivalent formulation
 for the force-constants matrix in linear response theory:
\begin{eqnarray}
&&C_{IJ}(\omega,T) 
=2 \sum_{\k i, \k^{\prime} j}^{N_{k}(T)}  \frac{f_{\k i}(T) -f_{\k^{\prime}
    j}(T)}{\eps_{\k i}-\eps_{\k^{\prime} j}+\omega+i\eta}
\nonumber \\
&\times&\langle  \psi_{\k^{\prime} j}|{\delta \vscf(\r,\omega)\over\delta
  \U_{I}}|\psi_{\k i}  \rangle 
\langle \psi_{\k i} |{\delta \vscf(\r,\omega) \over \delta \U_{J} }|\psi_{\k^{\prime} j}\rangle  \nonumber \\
&+& \int d\r \, n_0(\r)
{\delta^2 V_\mathrm{ext}(\r) \over \delta \U_{I}
\delta \U_{J}}  \nonumber \\
&-&\int\int n_{J}^{1}(\r,\omega,T)
K(\r,\r')
n_{I}^{1} (\r^{\prime},\omega,T)
d\r d\r^{\prime}.\nonumber \\
\label{eq:DE_DRDR} 
\end{eqnarray}

In Eq. \ref{eq:DE_DRDR} the term including the second derivative of the
external potential is real whereas all the other terms are complex.
The advantage of Eq. \ref{eq:DE_DRDR} is that it allows us to introduce a 
variational formulation of the
 force-constants matrix as it will be shown in the following.

\subsection{Variational formulation of the force-constants. \label{sec:varph}}
We introduce  the following force-constants functional $F_{IJ}$:
\begin{eqnarray}
&&F_{IJ}[\rho(\r),\rho^{\prime}(\r),\omega,T]=\nonumber \\
&=&2 \sum_{\k i,\k^{\prime} j}^{N_{k}(T)} \frac{f_{\k i}(T) -f_{\k^{\prime}j}(T)}
{\eps_{\k i}-\eps_{\k^{\prime} j}-\omega+i\eta }\nonumber \\
&\times&\langle \psi_{\k^{\prime} j}|
{\delta V_\mathrm{ext}(\r) \over \delta \U_{I}}+
\int K(\r,\r')
 \rho(\r')d\r'
|\psi_{\k i}\rangle \nonumber \\ 
&\times&\langle \psi_{\k i} |
{\delta V_\mathrm{ext}(\r) \over \delta \U_{J}}+
\int K(\r,\r')
 \rho^{\prime}(\r^{\prime})d\r^{\prime}
|\psi_{\k^{\prime} j}\rangle  \nonumber \\
&+& \int d\r \, n_0(\r)
{\delta^2 V_\mathrm{ext}(\r) \over \delta \U_{I}
\delta \U_{J}}  \nonumber \\
&-&\int\int \rho(\r)
K(\r,\r')
\rho^{\prime} (\r^{\prime})
d\r d\r^{\prime}.
\label{eq:functional}
\end{eqnarray}
With this definition the force-constants matrix reads as:
\begin{eqnarray}
C_{IJ}(\omega,T)  =F_{IJ}[n^{1}_I(\r,\omega,T),n^{1}_J(\r,\omega,T),\omega,T]
\label{eq:defCFij}
\end{eqnarray}

It is straightforward to note that the force-constant functional is quadratic 
with respect to $\rho$, namely:
\begin{eqnarray}
\left. {\delta F_{IJ}[\rho(\r),\rho^{\prime}(\r),\omega,T] \over \delta \rho(\r)}
\right|_{\rho(\r)=n^{1}_I(\r,\omega,T),\rho^{\prime}(\r)=n^{1}_J(\r,\omega,T)}=0.\nonumber \\
\label{eq:Var_ph_omega}
\end{eqnarray}
The same relation holds upon derivation with respect to $\rho'$.

The consequence of Eq. \ref{eq:Var_ph_omega} is that $C_{IJ}(\omega,T)$  is an 
extremal point and 
that a given error on $n^{1}_I(\r,\omega,T)$ or on $n^{1}_J(\r,\omega,T)$ 
affects the functional 
and the phonon frequencies only at second order. 

Our functional formulation is related to that used for dielectric tensors by 
Gonze {\it et al.} in Ref.  \onlinecite{GonzePRL92}. The main difference is that, 
while the functional formulation of Ref. \onlinecite{GonzePRL92} is variational with
respect to the first-order perturbation of the wave-function, our formulation is
variational with respect to the first-order perturbation of the electronic charge density.

\subsection{Approximated force-constants functional. \label{sec:appr_fun}}

At this point we can exploit the results of the preceding section in order to develop a method to accurately
calculate phonon and electron-phonon properties.
Within density functional theory the most precise force-constant matrix 
 $C_{IJ}(\omega,T)$ is
obtained when $T$ coincides with the physical temperature $T_0$ ($e.g.$ room
temperature) of the system. However, in a metal, this temperature is prohibitively small and the 
self-consistent calculation of $C_{IJ}(\omega,T_0)$ would be computationally unfeasible as it requires a large number of k-points in order to properly 
converge the summation in Eq. \ref{eq:DE_DRDR}. 
In addition, the self-consistent
calculation of the imaginary part of $n^{1}_I(\r,\omega,T)$ and 
of $n^{1}_J(\r,\omega,T)$ at finite $\omega\ne 0$
requires to increase further the number of k-points  with respect to standard  
static linear-response calculations of $n^{1}_I(\r,0,T)$.

A remedy to these problems is to define an approximated force constants matrix, 
$\tilde{C}_{IJ}$, as:
\begin{eqnarray}
\tilde{{ C}}_{IJ}(\omega,T_0)=
F_{IJ}[n^{1}_I(\r,0,T_{\rm ph}),n^{1}_J(\r,0,T_{\rm ph}),\omega,T_0]\nonumber \\
\label{eq:CA_best}
\end{eqnarray}
that is, $(i)$ the frequency dependence on $n^{1}_I(\r,\omega,T_{\rm ph})$ 
is neglected and its static 
limit ($\omega=0$) is considered and 
(ii) $n^{1}_{I}$ and $n^{1}_{J}$ are  
calculated at the temperature $T_{\rm ph}$ instead of $T_{0}$. 
The quantity T$_{\rm ph}$ is the electronic temperature commonly used in a 
self-consistent linear
response calculation. Usually T$_{\rm ph}$ is much larger then the physical temperature
$T_0$, ($e.g.$ $T_{\rm ph}\approx 4000$ K)\cite{comment}. 
The main consequence of the variational property of the functional $F_{IJ}$ is that
it allows us to compute phonon frequencies from an approximate force-constants
matrix $\tilde{{ C}}_{IJ}(\omega,T_0)$ in a {\it non self-consistent} way
and performing an error that is quadratic in $|n^{1}_I(\r,\omega,T_0) - n^{1}_I(\r,0,T_{\rm ph})|$.

Thus, a time consuming self-consistent calculation of the non-adiabatic force-constants $C_{IJ}(\omega, T_{0})$ has been replaced by an approximate 
non self-consistent one, $\tilde{C}_{IJ}(\omega, T_0)$, with an error that is negligible for the phonon frequencies, as it will be demonstrated in the applications considered
in this work.

We now first add and subtract $C_{IJ}(0, T_{\rm ph})$ (the standard linear-response adiabatic self-consistent force constants) in the left member of 
Eq. \ref{eq:CA_best}, and then perform a Fourier transform to obtain:
\begin{eqnarray}
&&\tilde{{ C}}_{sr}(\q,\omega,T_0)=\Pi_{sr}(\q,\omega,T_0)+
\nonumber \\
&&\,\,\,\,\,\,\,\,\,\,\,\,\,\,\,\,\,\,+C_{sr}(\q,0,T_{\rm ph})
\label{eq:CAdef2}
\end{eqnarray}
where
\begin{eqnarray}
&&\Pi_{sr}(\q,\omega,T_0)=\nonumber \\
&&=\frac{2}{N_k(T_0)} \sum_{\k i j}^{N_{k}(T_0)}
\frac{f_{\k i}(T_0)-f_{\k+\q j}(T_0)}
{\eps_{\k i}-\eps_{\k+\q j}+\omega+i\eta }\nonumber \\
&&\times  \,\,\,\,\,\,\,\,\,{\bf d}_{ij}^{s}(\k,\k+\q){\bf d}_{ji}^{r}(\k+\q,\k) \nonumber \\
&&-\frac{2}{N_k(T_{\rm ph})} \sum_{\k i j}^{N_{k}(T_{\rm ph})}
\frac{f_{\k i}(T_{\rm ph})-f_{\k+\q j}(T_{\rm ph})}
{\eps_{\k i}-\eps_{\k+\q j} }\nonumber \\
&&\times \,\,\,\,\,\,\,\,\,{\bf d}_{ij}^{s}(\k,\k+\q){\bf d}_{ji}^{r}(\k+\q,\k)
\label{eq:Pidef}
\end{eqnarray}
with the following definition of the deformation potential matrix element
\begin{eqnarray}
{\bf d}_{mn}^{s}(\k+\q,\k)&=&
\langle \k+\q m|\frac{\delta v_{\rm SCF}}{\delta {\bf u}_{\q s}}|\k n\rangle
\label{eq:gxq}
\end{eqnarray}
and where $|\k n\rangle$ is the periodic part of the Bloch  wavefunction, i.e. 
$|\psi_{\k n}\rangle=e^{i\k\cdot\r}|\k n\rangle/\sqrt{N_k}$, and
${\bf u}_{\q s}$ is the Fourier transform of the phonon
displacement $\u_{L s}$. The integration in Eq. \ref{eq:gxq} is understood to be
on the unit cell. The quantity
$ v_{\rm SCF}$ is the periodic part of the static self-consistent potential, namely
$\vscf(\r)=v_{\rm SCF} (\r) e^{i\q\cdot\r}$. This static self-consistent potential
is calculated using standard linear response at temperature $T_{\rm ph}$.

The calculation of $\Pi_{sr}(\q,\omega,T_0)$ requires
the knowledge of band energies and eigenfunctions on a denser 
$N_{k}(T_0)$ k-point grid using an electronic temperature $T_0$
only in an energy window of width $\sim\max(T_{\rm ph},\omega)$ around the Fermi level. Moreover it does
not require to recalculate the derivative of the self-consistent potential. As such
it is much less time consuming then an linear-response self-consistent
calculation to obtain $C_{rs}(\q,0,T_{0})$. 
The non-adiabatic phonon frequencies are calculated (in the clean limit)
non self-consistently from the Hermitian  part of $\tilde{C}_{sr}(\q,\omega,T_0)$
(see Eq. \ref{eq:redynmat}).

The expression of the adiabatic force-constants at the physical temperature $T_0$ is obtained setting $\omega=0$ in the 
Eq. \ref{eq:CAdef2}:  
\begin{eqnarray}
&&\tilde{C}_{sr}(\q,\omega=0,T_0)=\Pi_{sr}(\q,0,T_0)+\nonumber \\
&&\,\,\,\,\,\,\,\,\,\,\,\,\,\,\,\,\,\,+C_{sr}(\q,0, T_{\rm ph})
\label{eq:cadiab}
\end{eqnarray}
In this case the force constants are 
Hermitian  and the error in the approximated force constants 
is quadratic in 
$|n^{1}_I(\r,0,T_0) - n^{1}_I(\r,0,T_{\rm ph})|$.

The advantage of the present procedure is that the linear-response self-consistent
calculation is performed with a small number of k-points $N_{k}(T_{\rm ph})$ whereas
the low temperature force constants are obtained with a non self-consistent
calculation over much denser $N_k(T_0)$ k-points mesh. This approach requires,
as in a conventional electron-phonon coupling calculation, the knowledge of the
wavefunctions in a dense $N_k(T_0)$ k-points grid and in an energy window
of the order of the maximum between $T_{\rm ph}$ and $\omega$ around the Fermi level, but does not
require the self-consistent linear-response calculation of the derivative of 
the Kohn-Sham potential with respect to an atomic displacement.
As such the procedure is substantially less time consuming.

\subsection{Phonon frequencies interpolation over k-points at fixed
phonon-momentum: practical implementation\label{sec:pract1}}

The practical implementation of the above theoretical formulation for the 
 phonon frequencies proceeds as follows:

\begin{enumerate}
\item Perform a standard linear response calculation {\it for a given
      phonon momentum $\q$} 
 using a $N_{k}(T_{\rm ph})$ k-points mesh and smearing $T_{\rm ph}$
for the electronic integration  
to obtain ${ C}_{sr}(\q,0,T_{\rm ph})$.

\item In order to perform the second summation
in $\Pi_{sr}(\q,\omega,T_0)$ (Eq. \ref{eq:Pidef}),
calculate deformation potential matrix element
of Eq. \ref{eq:gxq} on the electron-momentum grid composed of
$N_{k}(T_{\rm ph})$ k-points using a smearing  $T_{\rm ph}$ .

\item Generate wavefunctions  
$|\k n \rangle $ and energies $\eps_{\k n}$
on the  denser $N_{k}(T_0)$ electron-momentum k-points grid.

\item Perform a second non self-consistent 
 calculation of the deformation-potential matrix-element  
on the more dense $N_{k}(T_0)$ k-points grid using smearing $T_{0}$ and
$v_{\rm SCF}$ obtained on the $N_{k}(T_{\rm ph})$ grid 
to obtain the first summation in $\Pi_{sr}(\q,\omega,T_0)$.

\item Calculate 
$\tilde{{ C}}_{sr}(\q,\omega,T_{0})$ (or $\tilde{{ C}}_{sr}(\q,\omega=0, T_{0})$)
using Eq. \ref{eq:CAdef2} (or Eq. \ref{eq:cadiab}).

\item Diagonalize the dynamical matrix to obtain phonon frequencies 
and phonon eigenvectors.

\end{enumerate} 
 
The procedure illustrated in this section is used to obtain
{\it at a fixed phonon momentum $\q$} well converged
phonon frequencies with respect to electronic k-points and smearing. 
Then standard Fourier interpolation can be performed  
to obtain dynamical matrices throughout the BZ.
However, this procedure,  has still two main computational shortcomings.
First, the calculations of the wavefunctions and matrix elements (points 3 and 4)
become  cumbersome when very dense $N_{k}(T_0)$ k-points grid for the
electron-momentum are necessary 
to converge.
Second, in presence of phonon anomalies and not-too-smooth phonon dispersion a dense k-sampling of the phonon BZ is required.

An optimal solution to overcome these problems is represented by the use of maximally localized 
Wannier functions\cite{Marzari, MarzariDIS} as an alternative electron basis function.

The method and the implementation of the Wannier functions approach for the electron-phonon
interpolation  is presented in Sec.~\ref{sec:wannier}.

\subsection{Phonon-linewidth and  of the electron-phonon coupling \label{sec:interp_lambda}}

The imaginary part of the force-constants matrix is related to phonon-damping,
namely the energy-conserving decay of a phonon in particle-hole pairs.

The imaginary part of Eq.
\ref{eq:Cvscfvb} is:
\begin{eqnarray}
& &\mathrm{Im}(C_{IJ}(\omega,T) ) =2 \sum_{\k i, \k^{\prime} j}^{N_{k}(T)} (f_{\k i}(T) -f_{\k^{\prime} j}(T))\nonumber \\
& &\times\left(\pi
\langle \psi_{\k^{\prime} j}|\mathrm{Re}\left[\frac{\delta\vscf(\r,\omega)}{\delta \U_{I}}\right]|\psi_{\k i}\rangle
\langle\psi_{\k i}|\frac{\delta V_\mathrm{ext}(\r)}{ \delta \U_{J}} 
|\psi_{\k^{\prime} j}\rangle\right.\nonumber \\
& &\times
\delta(\eps_{\k i}-\eps_{\k^{\prime} j}+\omega)\nonumber \\
& &+
\langle \psi_{\k^{\prime} j}|\mathrm{Im}
\left[\frac{\delta\vscf(\r,\omega)}{\delta \U_{I}}\right]|\psi_{\k i}\rangle
\langle\psi_{\k i}|\frac{\delta V_\mathrm{ext}(\r) }{ \delta \U_{J}} 
|\psi_{\k^{\prime} j}\rangle\nonumber \\
&&\times\left.P\left[\frac{1}{\eps_{\k i}-\eps_{\k^{\prime} j}+\omega}\right]\right)
\label{eq:ImC}
\end{eqnarray}
where $P$ indicates the principal part.
To obtain Eq. \ref{eq:ImC} we have assume that the unperturbed Hamiltonian 
is time-reversal symmetric so that a real set of eigenfunctions exists.

This expression is exact but it has two disadvantages. First it requires the expensive calculation of 
the derivative of the self-consistent potential at finite frequency both in its real
and imaginary parts. Second the phonon-decay in electron-hole pairs is not manifest
in the term including the principal part. These problems persist if the imaginary
part of Eq. \ref{eq:DE_DRDR} is considered, since, in this functional, both
${\delta\vscf(\r,\omega)}/{\delta \U_{I}}$ and $n^{1}_I(\r,\omega,T)$ are complex quantities.
These two shortcomings are absent if the approximated force-constants matrix 
in Eq. \ref{eq:CA_best} is used. Indeed the imaginary part of Eq. \ref{eq:CA_best} is:
\begin{eqnarray}
& &\mathrm{Im}(\tilde C_{IJ}(\omega,T) ) =2\pi \sum_{\k i, \k^{\prime} j}^{N_{k}(T)} (f_{\k i}(T) -f_{\k^{\prime} j}(T))\nonumber \\
& &\times
\langle \psi_{\k^{\prime} j}|\frac{\delta\vscf(\r,0)}{\delta \U_{I}}|\psi_{\k i}\rangle
\langle\psi_{\k i}|\frac{\delta \vscf(\r,0)}{ \delta \U_{J}} 
|\psi_{\k^{\prime} j}\rangle\nonumber \\
& &\times
\delta(\eps_{\k i}-\eps_{\k^{\prime} j}+\omega)
\label{eq:ImCapp}
\end{eqnarray}
with an error quadratic in $|n^{1}_I(\r,\omega)-n^{1}_I(\r,0)|$.
Fourier transforming and replacing Eq. \ref{eq:ImCapp} in Eq. \ref{eq:lw_C}
 gives the well known equation
for the phonon linewidth $\gamma_{{ \q}\nu}$ full-width half-maximum (FWHM), that is:
\begin{eqnarray}
\gamma_{{ \q}\nu}^{\rm Fermi}&=&\frac{4\pi}{N_k(T)}
\sum_{{ \k},m,n}^{N_k(T)}|g_{nm}^{\nu}(\k, \k+ \q)|^2  \nonumber \\
&\times&\left(f_{{ \k}n} - f_{{ \k}+{ \q}m}\right)
 \delta(\epsilon_{{ \k}+{ \q}m}-\epsilon_{{ \k}n}-\omega_{{ \q}\nu}),
\label{eq:gamma_fermi} 
\end{eqnarray}
where 
$
g_{nm}^{\nu}(\k, \k+ \q)=\sum_{s}
{\bf e}^{s}_{\q\nu}\cdot\d_{mn}^{s}(\k+\q,\k)/
{\sqrt{2M_s\omega_{\q\nu}}}
$.

In absence of the electron-electron interaction a similar expression involving only
the derivative of the bare external potential is easily obtained using Fermi golden rule. 
In the presence of electron-electron
interaction our formalism justify the replacement of the derivative of the bare
external potential 
by the derivative of the static screened potential $\vscf(\r,0)$ in the Fermi golden rule, with a well controlled error. 

Under certain conditions illustrated in Refs. \onlinecite{Allen,CalandraPRB05},
the phonon linewidth can be reduced to:
\begin{eqnarray}
&&\tilde{\gamma}_{{ \q}\nu} = \frac{4\pi\omega_{\q \nu}}{N_k(T)}\sum_{{ \k},m,n}^{N_k(T)}|g_{nm}^{\nu}(\k, \k+ \q)|^2 \times \nonumber \\
&&\delta(\epsilon_{{ \k}n})\delta(\epsilon_{{ \k}+{ \q}m}-\epsilon_{{ \k}n}-\omega_{{ \q}\nu})
\label{eq:gamma_omega}
\end{eqnarray}
Finally, neglecting $\omega_{\q\nu}$ in Eq. \ref{eq:gamma_omega}, we obtain Allen formula
\cite{Allen} namely,
\begin{equation}
\label{eq:gamma_Allen}
\gamma_{{\bf q}\nu} = \frac{4\pi \omega_{{\q \nu}}}{N_k(T)} \sum_{{\bf k},n,m}^{N_k(T)} |g_{nm}^{\nu}(\k, \k+ \q)|^2 \delta(\varepsilon_{{\bf k}n}) \delta(\varepsilon_{{\bf k+q}m})
\end{equation}

The Allen-formula \cite{Allen,Grimvall} is widely used, and represents a good estimation of the phonon-linewidth due to 
electron-phonon effects,  in the absence of anharmonic effects and other scattering processes.
In addition, the Allen-formula relates the  phonon-linewidth, as measured by inelastic Neutron or X-ray measurements, 
to the electron-phonon coupling as:
\begin{equation}
\lambda_{{\bf q}\nu}= \frac{{\gamma}_{{\bf q}\nu}}{2\pi \omega_{{\bf q}\nu}^2{\cal N}_{\rm s}},
\label{eqAllen}
\end{equation}
where ${\cal N}_{\rm s}$ is the electronic density of states per spin at the Fermi energy.

With this definition, the isotropic Eliashberg-function is
\begin{equation}
\alpha^2F(\omega)=\frac{1}{2 N_{q}}\sum_{{\bf q}\nu} 
\lambda_{{\bf q}\nu} \omega_{{\bf q}\nu} \delta(\omega-\omega_{{\bf q}\nu} )
\label{eq:Eliashberg_function}
\end{equation}
where $N_q$ is the number of phonon wavevectors.
We also define the integrated electron-phonon coupling as
\begin{equation}
\lambda(\omega)=2\int_{0}^{\omega} \frac{\alpha^2F(\omega')}{\omega'} \,d\omega'.
\end{equation}

\section{Wannier Functions\label{sec:wannier}}

\subsection{Wannier interpolation of the electron-phonon matrix element\label{sec:wan}}

In this section we explain how to interpolate the electron-phonon matrix
element throughout the BZ using Wannier functions \cite{GiustinoPRB}.

Using standard first-principles methods a set of Bloch functions  
$\psi_{\k n}$ are generated 
 with $\k$ belonging 
{\it to a uniform $N_{k}^{w}$ k-points grid centered in $\Gamma$}.
A set of Wannier functions centered on site $\R$ are defined by the relation
\begin{eqnarray}\label{eq:bloch2wan}
|{\bf R} m \rangle =  \frac{1}{\sqrt{N_k^w}}
\sum_{\k n} e^{-i{\k}\cdot{\bf R}} U_{nm}({\k})|\psi_{\k n}\rangle 
\end{eqnarray}
A suitable transformation matrix, $U_{mn}(\k)$,  must be determined in the so-called Wannierization procedure. 
In this work we choose to work with Maximally Localized Wannier functions (MLWF)
\cite{Marzari,MarzariDIS}, although  other Wannierization schemes are possible.
In the MLWF case, the matrices $U_{nm}({\k})$ are obtained following
the prescription of Refs. \onlinecite{Marzari,MarzariDIS}, which guarantees the maximal space localization 
of the final Wannier functions. This last requirement will be essential in the spirit of interpolation of the electron-phonon matrix elements.

Specifically, if the band-structure is formed by a composite group
of bands and the number of Wannier functions is identical to the number of
bands of the composite group of bands, then the matrix $U_{mn}(\k)$
is a square matrix. On the contrary if the desired bands are entangled in a larger manifold,  
a preliminary disentanglement procedure from the manifold must be performed 
and then the square matrix $U_{mn}(\k)$ is obtained\cite{MarzariDIS}.

The deformation potential matrix-elements $\d_{mn}^{s}(\k+\q,\k)$
(see Eq. \ref{eq:gxq}) are calculated using linear response
scheme. In this calculation particular care is needed for
the deformation potential at zone center, namely $\d_{mn}^{s}(\k,\k)$, as
explained in sec. \ref{sec:efshift}.

It is crucial to note that the periodic parts  $|\k n\rangle$ and
$|\k+\q m\rangle$ 
entering in $\d_{mn}^{s}(\k+\q,\k)$
have to be {\it exactly the same wavefunctions used for the
Wannierization procedure}. If this is not the case,  spurious (unphysical) phases appear 
in the $|\k n \rangle $'s ($i.e.$ a phase added by the 
diagonalization routine or other computational reasons) and the localization properties 
of the Wannier functions in real space are completely lost. 
A way to enforce this condition is to use a uniform grid centered at
$\Gamma$ so that $\k+\q=\k'+\G$ with $\k'$ still belonging to the original grid. 
In this way  $|\k+\q n \rangle$  can
be obtained from $|\k' n \rangle $  simply multiplying by a phase determined by the $\G$-vector translation 
and no additional phases occur.

Exploiting translational invariance,
the deformation-potential matrix-element in the Wannier function basis is
obtained by  Fourier transform as,
\begin{eqnarray}\label{eq:grr}
& &\d_{mn}^{s}(\R, \R_{L})=\langle {\bf 0}m |\frac{\delta \vscf}{\delta \u_{s, L}}|{\bf R}n\rangle \nonumber \\
& &= \frac{1}{N_{k}^{w}}\sum_{\k,\q}^{N_{k}^{w}}\sum_{m',n'}
e^{-i{\bf k}\cdot \R+i\q\cdot\R_L} \nonumber \\
& &\times
U_{m m'}^{*}(\k+\q)\d^{s}_{m'n'}(\k+\q,\k) U_{n'n}(\k)
\nonumber \\
\end{eqnarray}
where $\R$ and $\R_L$ belong to a $N_k^w$ real-space supercell.
At this point it is important to underline that {\it the grid of $N_k^{w}$ $k-$points
on which the Wannierization process has been carried out has to be exactly the same
grid for the phonon-momentum 
on which the dynamical matrices have been computed}.
One linear response calculation for each k-point in the phonon irreducible BZ (IBZ)
of the $N_k^w$ electron k-points grid has to be carried out. If this constraint is
not respected, the localization properties of the electron-phonon matrix element
in real space is not guaranteed.

Inverting Eq. \ref{eq:bloch2wan}, we obtain
\begin{eqnarray}\label{eq:wan2bloch}
|\psi_{\k n}\rangle = \frac{1}{\sqrt{N_{k}^{w}}}\sum_{\R}\sum_{m} e^{i\k\R} U_{nm}^{*}(\k)
|\R m\rangle
\end{eqnarray}
Noting that
$ \d_{mn}^{s}(\k+\q,\k)=\frac{1}{N_{k}^{w}}
\langle \psi_{\k+\q m}|\frac{\delta \vscf}{\delta \u_{\q s}}|\psi_{\k n}\rangle$, and
using Eq. \ref{eq:wan2bloch} one obtains:
\begin{eqnarray}\label{eq:wank2r}
 & &\d_{mn}^{s}(\k+ \q, \k)=
\frac{1}{(N_k^w)^2}\sum_{L}\sum_{\R}\sum_{m'n'}e^{i \k\cdot{\bf R}+i \q\cdot{\bf R}_L}\nonumber \\
& & U_{m'm}(\k+\q){\bf d}_{m'n'}^{s}(\R, \R_L) U_{nn'}^{*}(\k)
\nonumber \\
\end{eqnarray}
where $\R$ and $\R_L$ belong to a $N_k^w$ real-space supercell.
Now, if $\d_{mn}^{s}(\R, \R_L)$ is localized inside the 
$N_k^w$ real-space supercell then
the interaction between different $N_k^w$ real-space supercells 
can be neglected and $d_{mn}^{s}(\k+\q,\k)$ can be obtained
from Eq. \ref{eq:wank2r} via a slow Fourier transform. In
practice this means that in Eq. \ref{eq:wank2r}
now $\k$ and $\q$ are any k-points in the BZ.
As a result of the interpolation, $\d_{mn}^{s}(\k+\q,\k)$
can be used to calculate any physical property as in a simple tight-binding scheme.

\subsection{Deformation potential matrix elements of optical zone-center phonons  
\label{sec:efshift}}

In this section we discuss the peculiarities related to the calculation of the electron-phonon matrix elements for optical zone center phonons ($\bf q = 0$). 
The periodic part of the self-consistent potential, induced by the phonon displacement at wavelength $\bf q$ of the atom $s$  can be decomposed in the Coulomb (Cl) potential (the sum of the bare and Hartree potentials) and exchange-correlation (XC) contributions:
\begin{equation}
\frac{\delta v_{\rm SCF} ({\bf r})}{\delta {\bf u}_{{\bf q}{s}}}=
\frac{\delta v_{\rm XC} ({\bf r})}{\delta {\bf u}_{{\bf q}{s}}}+
\frac{\delta v_{\rm Cl} ({\bf r})}{\delta {\bf u}_{{\bf q}{s}}}
\end{equation}
The integral of the Coulomb contribution over the unit cell volume 
$\Omega$ is:
\begin{equation}
\Delta_{{\bf q}{s}}=\frac{1}{\Omega}\int d^3r 
\frac {\delta v_{\rm Cl} ({\bf r})}
{\delta {\bf u}_{{\bf q}{s}}}.
\end{equation}
In DFT linear response codes, such as QUANTUM-ESPRESSO\cite{QE},  the phonon
calculation with $\bf q \ne 0$ and $\bf q =0$ are treated with two different approaches.\cite{deGironcoli95,pwscfRMP}

At $\bf q \ne 0$, the calculation is performed within the gran-canonical 
ensemble, with a constant electron chemical-potential (Fermi level) $\epsilon_{\rm F}$. 
In a metal, the limit
\begin{equation}
\lim_{\q \rightarrow {\bf 0}} \Delta_{{\bf q}{s}}=\Delta_{{0^+}{s}}
\end{equation}
is well defined and independent on the direction of $\bf q$. 
In general one has $\Delta_{0^+s}\ne  0$.

At $\bf q = 0$,  the calculation is performed in the canonical ensemble,
with a constant number of electrons. In this calculation the variation of the average Coulomb potential
is conventionally  set to zero, namely:
\begin{equation}
\Delta_{{\bf 0}{s}}=0
\end{equation}
To keep constant the number of electrons the derivative of the Fermi energy with respect to the phonon displacement $\delta \epsilon_{\rm F}/\delta_{{\bf u}{\bf 0}{s}}$ can be different from zero. In particular\cite{deGironcoli95,pwscfRMP} it must be:
\begin{equation}
\frac{\delta \epsilon_{\rm F}}{\delta \u_{\bf 0 s}}=-\Delta_{0^+s}
\end{equation}
In the QUANTUM-ESPRESSO implementation \cite{QE} 
the  ${\delta v_{\rm SCF} ({\bf r})}/{\delta {\bf u}_{{\bf q}{s}}}$ stored-on-disk is discontinuous at $\bf q =0$, 
since, in this case, it does not include (as it should) the contribution from
the average Coulomb potential.
As a consequence the electron phonon matrix elements computed with ${\delta v_{\rm SCF} ({\bf r})}/{\delta {\bf u}_{{\bf 0}{s}}}$ are incorrect. 
Moreover such discontinuity deteriorates
the localization properties of the electron-phonon
matrix elements in real space.
The problem is easily solved by redefining the self-consistent potential to be used in the calculation of the deformation-potential matrix-elements for $\bf q =0$ as:
\begin{equation}
\frac{\delta \tilde v_{\rm SCF} ({\bf r})}{\delta {\bf u}_{{\bf 0}{s}}}=
\frac{\delta v_{\rm SCF} ({\bf r})}{\delta {\bf u}_{{\bf 0}{s}}}-
\frac{\delta \epsilon_{\rm F}} {\delta {\bf u}_{{\bf 0}{s}}} .
\end{equation}
Note that since ${\delta \epsilon_{\rm F}}/\delta{\bf u}_{{\bf 0}s}$
transforms under symmetry operation as a force, 
it is different from zero only for the atoms for which the internal 
coordinates are not fixed by symmetry.
For this reason, $\delta\epsilon_{\rm F}/\delta{\bf u}_{{\bf 0}s}$
is finite 
in CaC$_6$, while it vanishes in MgB$_2$.

We remark that the same problem occurs when 
a frozen-phonon calculation of 
$\frac{\delta v_{\rm SCF} ({\bf r})} {\delta{\bf u}_{ {\bf 0} s}} $
is carried out as in all the electronic-structure codes the 
average Coulomb potential is conventionally set to zero.

\subsection{Wannier interpolation of the dynamical matrix at fixed 
phonon-momentum.\label{sec:wanfixq}}

Steps $4,5,6$ of section \ref{sec:pract1} can be repeated using Wannier 
functions to interpolate at fixed phonon momentum $\q$ the quantity
 $\Pi_{sr}(\q,\omega,T_0)$
to obtain Eq. \ref{eq:Pidef}. 
The advantage of the Wannier-interpolation-based
strategy with respect to the procedure explained in sec. \ref{sec:pract1} is that 
the time needed to calculate the matrix element is now negligible as 
there is no need to obtain, from first-principles, wavefunctions and energies on the dense
 $N_k(T_0)$ k-point grid.
Still it is necessary to perform Fourier interpolation to obtain 
dynamical matrices throughout the full BZ.

\subsection{Wannier interpolation of the dynamical matrix. \label{sec:interp_all}}

In what follows we outline a strategy to obtain dynamical matrices
throughout the BZ using both Fourier interpolation and Wannier functions.
The method solves the main problem related to the presence of Kohn-anomalies, originating 
from long range interactions in the dynamical matrix.
The idea is to separate in the force-constant matrix obtained
in Eq. \ref{eq:CAdef2} in the short and long range components. The long range 
force constants are associated to Kohn anomalies driven by Fermi surface
nesting, and, as such, they cannot be easily Fourier interpolated and require an accurate sampling 
of the Fermi surface.
This last part will be treated using the Wannier interpolation scheme explained in
section \ref{sec:wan}. On the contrary short range force constants produce smooth phonon dispersions and can be safely Fourier interpolated and do not require accurate sampling of the Fermi surface. 
Treating separately the two length scales permits  to obtain converged phonon frequencies with respect to k-point sampling  and electronic temperature, 
everywhere  in the Brillouin zone.

This procedure is similar to the treatment of long-range dipolar forces 
due to the Born effective charges in polar
semiconductors\cite{pwscfRMP,GonzePhonon}. 
In this case, these long range forces generate a non-analytical
behavior of the dynamical matrix at zone center. 
While Fourier interpolation of the dynamical matrices including the non-analytical
terms is impossible, it becomes possible if the long-range forces are subtracted
to obtain short-range force-constants.
The non-analytical part subtracted to 
perform Fourier interpolation is added again after Fourier interpolation.

In this spirit, we rewrite Eq. \ref{eq:CAdef2} as,
\begin{eqnarray}
&&\tilde{C}_{sr}(\q,\omega,T_0)=\Pi_{sr}(\q,\omega,T_0)\nonumber \\
&& -\Pi_{sr}(\q,0,T_{\infty})+\tilde{C}_{sr}(\q,0,T_{\infty})
\label{eq:Ca_vs_Cainf} 
\end{eqnarray}
where 
\begin{eqnarray}
\tilde{C}_{sr}(\q,0,T_{\infty})&=&C_{sr}(\q,0,T_{\rm ph})+\nonumber \\
&+&\Pi_{sr}(\q,0,T_{\infty})
\label{eq:C_HT}
\end{eqnarray}
and $T_{\infty}$ is an electronic temperature large enough in 
order to have only short range force constants named 
$\tilde{C}_{sr}(\q,0,T_{\infty})$,
and no Kohn anomalies
in the corresponding phonon branches.
As a consequence, Fourier
interpolation can be applied to $\tilde{C}_{sr}(\q,0,T_{\infty})$.

In practical implementation the procedure is the following:
\begin{enumerate}
\item{Obtain $C_{sr}(\q,0,T_{\rm ph})$ 
from  self-consistent
linear-response phonon calculations. Such force constants are
evaluated using a $N_{k}^{w}$ k-point grid for the phonon momentum
and an $N_{k}(T_{\rm ph})$ k-point grid for the electron momentum.}  
\item{Calculate $\Pi_{sr}(\q,0,T_{\infty})$ using the Wannier functions 
as described in section \ref{sec:wanfixq}. }
\item{Use Eq. \ref{eq:C_HT} to  compute 
$\tilde{C}_{sr}(\q,0,T_{\infty})$ on the
$N_{k}^{w}$ phonon momentum grid.}
\item{Fourier interpolate $\tilde{C}_{sr}(\q,0,T_{\infty})$
to obtain  dynamical matrices at any desired phonon
momentum at temperature $T_{\infty}$.}
\item{Obtain $\tilde{C}_{sr}(\q,\omega,T_0)$ in Eq.\ref{eq:Ca_vs_Cainf} 
by Wannier interpolation 
of $\Pi_{sr}(\q,\omega,T_0)-\Pi_{sr}(\q,0,T_{\infty})$ at any desired
phonon momentum.}
\item{Diagonalize the dynamical matrix associated to 
$\tilde{C}_{sr}(\q,\omega,T_0,T_{\rm ph})$
to obtain adiabatic, $\omega=0$, or non-adiabatic (in the clean limit) $\omega=\omega_{\q\nu}$,  
phonon frequencies. These phonon frequencies 
include Kohn-anomalies
as long range terms are properly treated by the Wannier-functions approach.}

\end{enumerate}

\section{Applications\label{sec:appli}}
We applied the theoretical and computational framework developed in the preceding sections to two well known superconductors, namely magnesium diboride (MgB$_2$) and Calcium intercalated graphite (CaC$_6$). In these systems  phonon, electron-phonon and superconducting properties are extensively investigated from both theoretical and experimental point of view and a detailed comparison can me made.
At the same time, they represent non-trivial examples, where convergence problems in first-principles approaches 
are relevant for the calculation of phonon and superconducting properties.

\subsection{MgB$_2$\label{mgb2}}

Magnesium diboride (MgB$_2$) is, without doubts, one of the most studied and investigated materials, since 2001.
The discovery of superconductivity at intermediate temperatures (40 K)\cite{akimitsu},  in this compound, represented an unexpected surprise in the scientific community.
Easily available, 
a huge amount of experimental characterizations  were immediately possible\cite{review}. 
On the other hand,  MgB$_2$ being composed of light elements in an hexagonal unit cell with only three atoms, allowed a first-principles computational approach for the theoretical description and understanding of its normal and superconducting properties.
Density Functional Theory and linear response techniques, revealed an extraordinary high  electron-phonon coupling between $sp^2$ bonding and anti-bonding boron orbitals, with  a particular in-plane mode ($E_{2g}$) of the hexagonal boron sheet\cite{kortus, pickett, andersen}. 
Based on model calculations of the electron-phonon coupling parameter ($\lambda\simeq1$
in Ref. \onlinecite{pickett}), 
the approximate  critical temperature was estimated to be around 40 K. 
However, a subsequent careful investigation of the problem, revealed that the first-principles calculated values of $\lambda$, 
were too low to explain this high critical temperature, unless unphysical low Coulomb pseudo-potentials were used\cite{mazin}. 
The solution of this inconsistency, has come with the discovery of two-band-superconductivity\cite{giubileo} in MgB$_2$. 
Two different gaps develop on the $\sigma$ and $\pi$ bands of the boron sheets\cite{mazin}.
This allows to gain new scattering channels to enhance T$_c$ still containing the average value of $\lambda$.
At the moment, MgB$_2$ is classified as a   two-band electron-phonon superconductor\cite{Choi, scdft}.

This continuous improvement of the theoretical explanation of the normal and superconducting properties of MgB$_2$ was in part due to some difficulties in the proper calculations of the electron-phonon coupling and of the phonon frequencies.
Soon after the discovery of superconductivity in MgB$_2$, the first theoretical papers (based on standard DFT codes) evidenced a relevant and anomalous  difficulty to properly "converge" the E$_{2g}$ phonon frequency and the electron-phonon coupling.
The reason is now clear: due to the two dimensional nature (almost cylindrical) 
of the $\sigma^{*}$ Fermi surfaces
electron-phonon coupling calculations  
require a careful integration of the Fermi surface is needed.
This translates, in practical calculations, with the necessity to use an extremely dense k- space integration for electronic and phononic degree of freedom\cite{matteo}. Although feasible in principle, it would require a huge amount of resources and computational time.

So, still after about ten years of calculations,  the fundamental question about the value of the electron-phonon coupling in MgB$_2$, remains unanswered. 
This information is fundamental in order to validate the electron-phonon origin of the superconductivity in MgB$_2$ and at the same time, include other, still unexplored pairing interactions.

The Wannnier interpolation scheme, developed in the present paper, of  phonon frequencies and electron-phonon  matrix elements, represents the main tool to solve this problem. It retains the first-principles character, but at the same time it is affordable from a computational point of view.

In MgB$_2$ the BZ-center phonon-frequency, measured by Raman spectroscopy, lies  between the adiabatic and clean-limit-non-adiabatic phonon-frequency \cite{Saitta} computed by DFT. Thus MgB$_2$ is in the intermediate regime 
(see Sec.~\ref{sec:adiabaticnonadiabatic} 
and the relaxation time estimated in Ref.~\onlinecite{Saitta}). 
Since the experimental frequency at BZ center is closer to the adiabatic than the clean-limit-non-adiabatic value,\cite{Saitta} in MgB$_2$ we will not consider non-adiabatic effects, and we will present only the adiabatic dispersion.

\subsubsection{Technical details}

\begin{figure}[h]
\includegraphics[width=0.8\columnwidth]{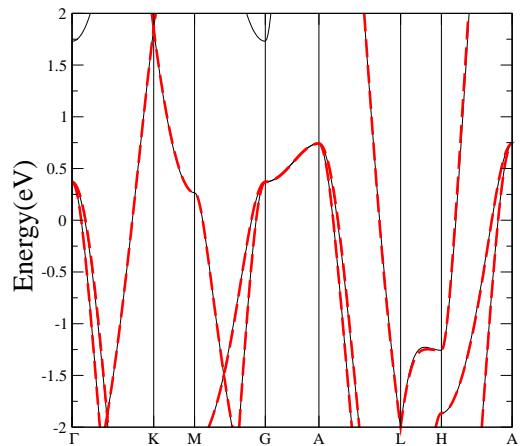}
 \caption{(color online) Wannier-interpolated (red-dashed) and first-principles-calculated bands (black-continuous) for MgB$_2$. The band-energies are plotted with respect to the Fermi level.}
\label{fig:bands_MgB2}
 \end{figure}

Wannier-interpolation of the MgB$_2$ band-structure is carried out using 
a $N_{k}^{w}=6\times 6\times 4$
k-points grid centered at $\Gamma$ 
using $5$ Wannier functions starting from random projections.
The Wannierized band structure is in excellent agreement with first-principles
calculation in a 1.5 eV region from the Fermi level, as shown in 
Fig. \ref{fig:bands_MgB2}.

The phonon dispersion and the electron-phonon matrix element are calculated 
using linear-response \cite{pwscfRMP,QE} on 
the irreducible part of the $N_{k}^{w}$ k-points grid centered at $\Gamma$.
The $\delta v_{\rm SCF}/\delta \u_{\q s}$ on the full grid are 
obtained applying symmetry operations of the crystal.
In the linear response calculation we perform
electronic integration using a $N_{k}=16\times 16\times 12$
$k-$point grid in the Brillouin zone and an Hermite-Gaussian smearing 
$T_{\rm ph}=0.025 $Ryd. 

The interpolation of phonon frequencies is performed along the 
strategy outlined in sec. \ref{sec:interp_all} and using 
$T_{\infty}=0.11$ Ryd 
to obtain smoothed ``high temperature''
phonon frequencies.

\subsubsection{Adiabatic phonon dispersion \label{sec:MgB2_ph}}

\begin{figure}[h]
\includegraphics[width=0.9\columnwidth]{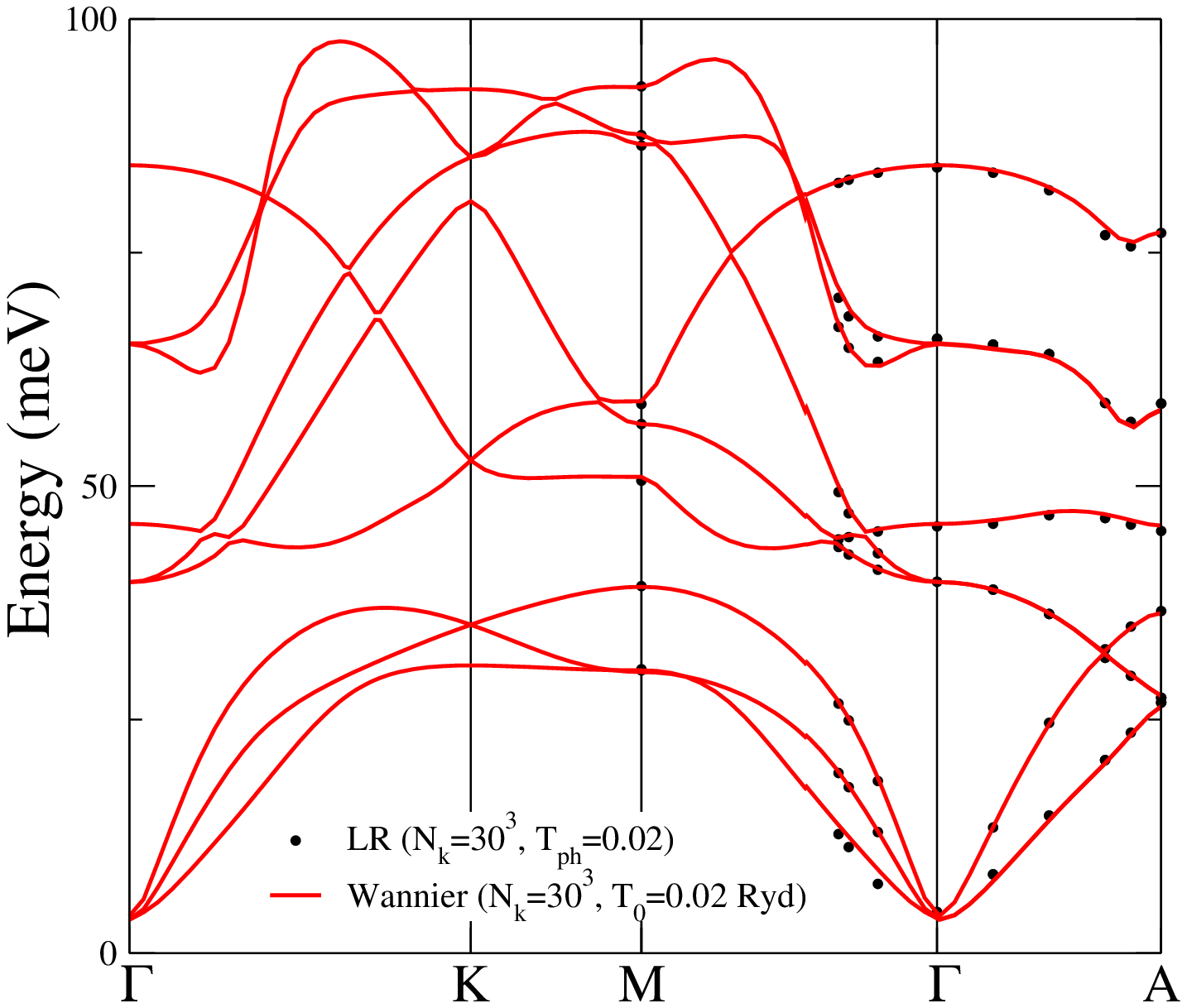}
\includegraphics[width=0.9\columnwidth]{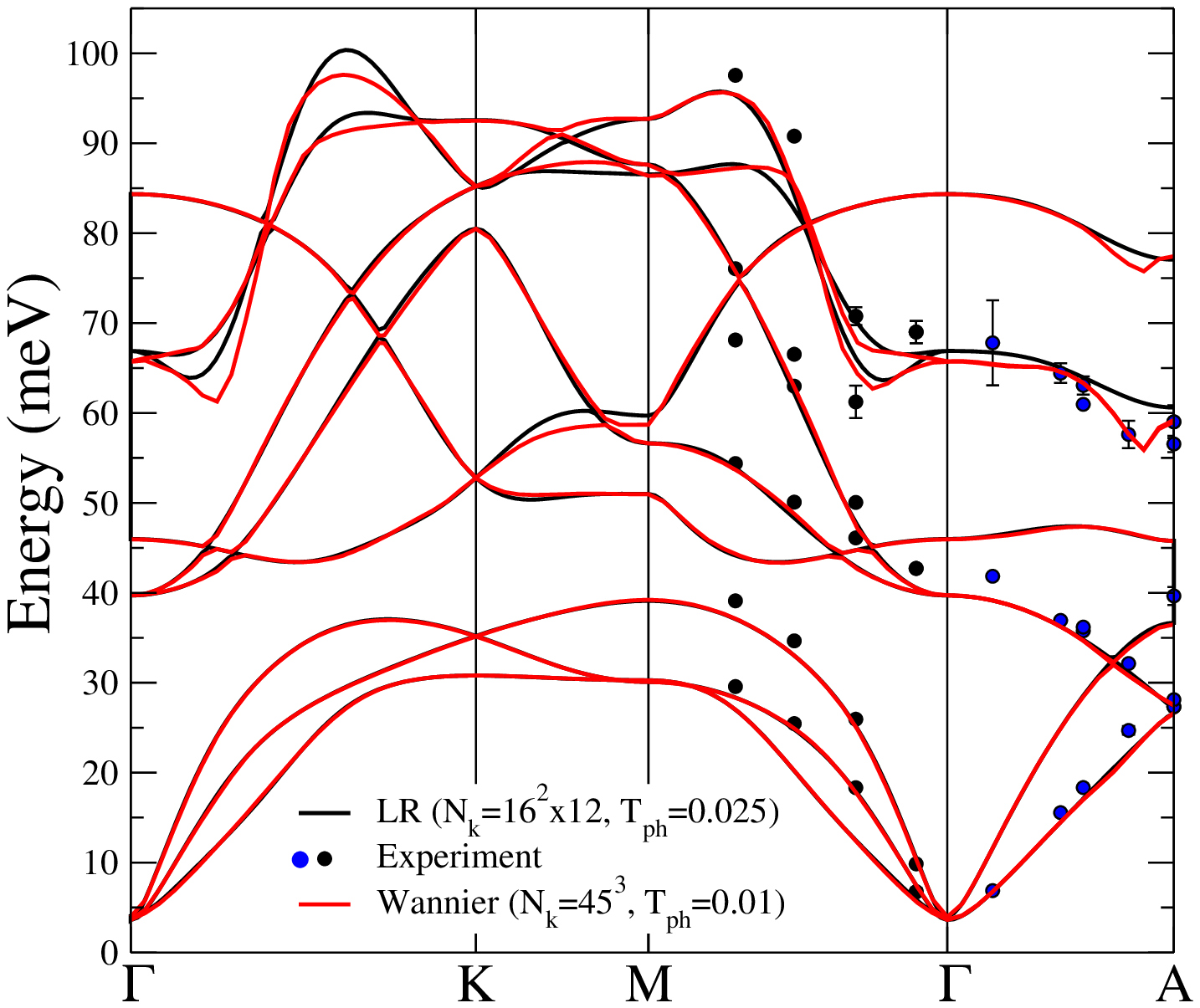}
 \caption{(color online) Top panel:
Wannier-interpolated adiabatic phonon dispersion  
(red line) compared to standard linear-response calculations
performed on selected points in the Brillouin zone (black dots).
Bottom-panel: Comparison between state-of-the-art calculated phonon
dispersion using Fourier interpolation (black line) and Wannier interpolation
(red line). Both interpolation schemes start from {\it the same} linear response
calculation. Experimental data in the bottom panel are from Ref. \onlinecite{Shukla} (blue circles) and from Ref. \onlinecite{Astuto} (black circles). 
Electronic temperatures are in Rydberg. The acoustic sum rule is not applied. 
Interpolated branches are obtained by connecting the points at which 
the interpolation has been carried out in the order of increasing energy.}
\label{fig:branchie_MgB2}
\end{figure}

The Wannier-interpolated adiabatic phonon dispersions on different 
k-point grids and smearings are compared to state-of-the-art electronic
structure calculations \cite{Kong,Bohnen,Choi} and experimental 
data \cite{Shukla,Astuto} in Fig. \ref{fig:branchie_MgB2}.
In the top panel of Fig. \ref{fig:branchie_MgB2} we compare the Wannier interpolated
phonon dispersion with $T_0=0.02$ Ryd and $N_{k}(T_0)=30^3$ with linear response 
calculations performed on selected points in the Brillouin zone 
using the same mesh and the same electronic temperature.
The error in the interpolation scheme is always smaller then $0.5$ meV.
In the bottom panel of Fig. \ref{fig:branchie_MgB2} we 
compare Wannier interpolated dispersion curves on N$_{k}(T_0)=45^{3}$ at $T_0=0.01$ Ryd
with Fourier interpolated linear response calculations from a 
$6\times 6\times 4$ phonon momentum grid. 
As can be seen anomalies are clearly found along $\Gamma A$ using our new method
but are completely missed by standard Fourier interpolation.
The main differences with respect to previous calculations
\cite{Kong,Bohnen,Choi} are :
\begin{enumerate}
\item{a prominent softening (Kohn anomaly) of the two E$_{2g}$ modes at
$\q\approx 0.25 {\Gamma{\rm  K}}$ is seen. The Kohn anomaly is also 
present along the $\Gamma$M high symmetry direction and in a circular region
around the zone center. Its origin is due to a in-plane nesting between the 
two $\sigma$ cylinders. This Kohn anomaly could be probably inferred
from previous calculations \cite{Kong,Bohnen,Shukla} on smaller k-points
grids although the softening
of the E$_{2g}$ modes in these works is weaker.}
\item{A second Kohn anomaly, absent in previous calculations 
\cite{Kong,Bohnen,Shukla}, 
is present on the E$_{2g}$ and B$_{1g}$ modes along the $k_z$ direction at 
$\q\approx 0.8 \Gamma$A. This Kohn anomaly
is due to nesting between different cylinders along $\Gamma$A. As it can be
seen by the comparison with previously available experimental data
in Fig. \ref{fig:branchie_MgB2}, its presence is confirmed
by inelastic X-ray scattering measurements of the phonon dispersion
\cite{Shukla,Astuto}. The data at $\q\approx 0.8 \Gamma$A are indeed in disagreement 
with previous phonon dispersion calculation\cite{Kong,Bohnen,Choi,Shukla,Astuto} even if 
this disagreement was overlooked in all previous publications. }
\item{Overall the Wannier-interpolated phonon dispersion is in better agreement
against experiments with respect to the linear-response calculated one. }
\end{enumerate}

These differences point out the need of having an ultra-dense k-point sampling
of the Brillouin zone not only for the electron-phonon coupling but also for the
phonon dispersion.

\subsubsection{Electron-phonon coupling \label{sec:MgB2_elph}}

Having interpolated the electron-phonon matrix elements and the
phonon frequencies it is possible
to calculate the phonon-linewidth and the electron-phonon coupling with a 
high degree of precision, eliminating source of errors coming from 
insufficient $k-$point sampling.
\begin{figure}[h]
\includegraphics[width=0.9\columnwidth]{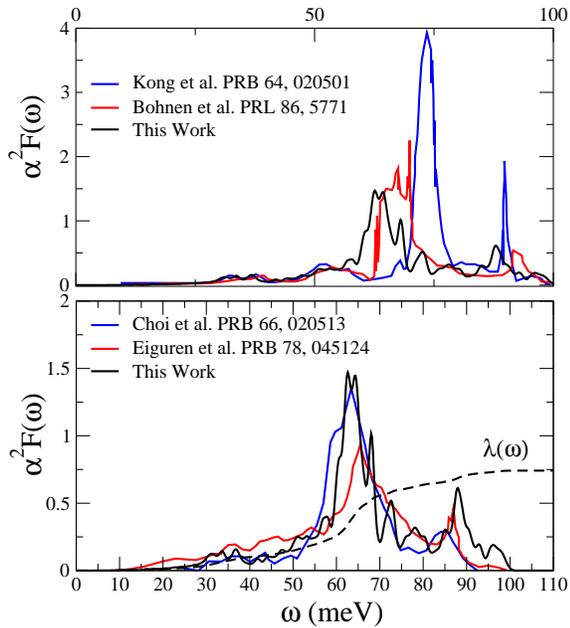}
 \caption{(color online) Comparison of our Wannier-interpolated MgB$_2$
Eliashberg function with previous calculations. Kong {\it et al.} \cite{Kong} 
used first-principles calculations and a particular treatment 
for $\lambda_{\q E_{2g} }$ for $\q$
along $\Gamma$A. Bohnen {\it et al.} \cite{Bohnen} used linear response
calculations while Choi {\it et al.}
\cite{Choi} performed frozen-phonon calculations. Finally Eiguren 
{\it et al.} \cite{Eiguren} 
used a different implementation of Wannier interpolation on the electron-phonon
matrix elements with
$40^3$ k-points grid for electron-momentum and $40^3$ k-points grid 
for phonon momentum, but did not interpolate phonon frequencies.
In this work we Wannier-interpolate both electron-phonon coupling
(using $80^3$ k-points grid for electron-momentum and $40^3$ k-points grid for
phonon momentum) {\it and} phonon frequencies
 (using $30^3$ k-points grid for electron-momentum and $40^3$ k-point
grid for phonon momentum).
The integrated $\lambda(\omega)$ from our work is also shown in the bottom panel.}
\label{fig:a2f_MgB2}
\end{figure}
In Fig. \ref{fig:a2f_MgB2} the Wannier-interpolated Eliashberg-function is compared 
with previous calculations done with different approaches. There are two main
improvements due 
to the better k-point sampling. 

First, several
previous calculations \cite{Kong, Bohnen, Eiguren}
on smaller k-point grids generally overestimate the position of the main peak 
in $\alpha^2 F(\omega)$ (centered around $\omega\approx 63$ meV in our work) due to
the coupling of $\sigma$-bands with the E$_{2g}$ phonon modes. This happens because
in previous calculations the Kohn anomalies of the E$_{2g}$ branches at 
$\q\approx 0.25\Gamma K$ was underestimated and because
the anomaly along $\Gamma A$ is missing. 
This crucial dependence of the
Eliashberg function on the E$_{2g}$ phonon frequency demonstrates the need of 
having both an accurate determination of $\lambda$ and of the phonon frequencies
to describe superconducting properties. 
 
Second, in our work there is substantial more weight then in previous works 
\cite{Kong,Eiguren} in the 90-100 meV region. This energy region is dominated
by the coupling of $\pi$ electrons to the E$_{2g}$  vibrations.
 
We now compare in more details our work with previous calculations.
In the work of Bohnen
{\it et al.} \cite{Bohnen} 
all the peaks are up-shifted not only by the limited k-points sampling
but also by the choice of using the DFT-minimized 
crystal structure and not the experimental one.
Kong and coworkers \cite{Kong} have calculated $\lambda_{\q \nu}$ and $\omega_{\q \nu}$ on
a $6\times 6\times 6$ phonon-momentum grid. 
Then for $\q$ of this grid belonging to the $\Gamma$A direction
the corresponding $\lambda_{\q \nu}$ values were replaced with some of the
$12\times 12\times 6$ k-point grid (see Ref. \onlinecite{Kong} for more details).
Fig. \ref{fig:a2f_MgB2} shows that this approximation strongly overestimates 
the energy of the peak due to coupling between $\sigma$ electronic states and 
the E$_{2g}$ mode. Indeed the average electron-phonon coupling is much larger
than the majority of the calculations present in literature (see Table \ref{tab:elph}).
Choi {\it et al.} \cite{Choi} used the LDA-minimized lattice structure 
but surprisingly obtained {\it softer} E$_{2g}$ phonon frequencies at $\Gamma$ then
all other works using experimental lattice parameters. 
In Ref. \onlinecite{Choi} 
the electron-phonon coupling is calculated using an interpolation
scheme that does not assure the localization properties of the electron-phonon
matrix element in real space. Nevertheless the position of the main peak in 
$\alpha^{2}F(\omega)$ is similar to what we obtain. 
On the contrary the secondary peak related to coupling between $\pi$
states and the E$_{2g}$ mode is at too low energy.  
Finally, Eiguren {\it et al.} \cite{Eiguren} used a Wannier interpolation scheme 
that requires an explicit calculation of the wavefunctions for any
phonon momentum. The interpolation scheme was used to calculate the average 
electron-phonon coupling 
but not to interpolate phonon frequencies as they were obtained
by Fourier interpolation.
As a consequence the phonon dispersion does not present the two observed
Kohn anomalies and the main peak of the $\alpha^{2}F(\omega)$ is at higher energies 
respect to our work. This difference reflects the need of having an accurate
determination of the phonon frequency in order to have converged Eliashberg 
functions. This appears in the lower phonon frequency logarithmic average
obtained in our work with respect to previous calculations as shown in table
\ref{tab:elph}. 
 \begin{table}
\begin{ruledtabular}
\begin{tabular}{l|c|c|c|c}
{\rm Reference}    & $\lambda$  & $\omega_{\rm log}$(meV) & $N_{k}$ & $N_{q}$ \\ \hline 
Kong {\it et al.} [\onlinecite{Kong}] & $0.87$  &  $62.0$ & $12^2\times 6$ & $6^3$ \\
Bohnen {\it et al.} [\onlinecite{Bohnen}] & $0.73$ & $60.9$ & $36^3$ & $6^3$ \\
Choi {\it et al.} [\onlinecite{Choi}] & $0.73$ & $62.7$ &  $18^{2}\times 12$ & $18^{2}\times 12$ \\
Eiguren {\it et al.} [\onlinecite{Eiguren}] & $0.776$ &   & $40^3$ &$10^3$ \\
This work & $0.741$ & $58.8$  & $80^3$ &$20^3$ \\
\end{tabular}
\end{ruledtabular}
 \caption{Comparison between different electron-phonon coupling calculations in MgB$_2$ using 
different first principles methods. The label N$_{k}$ (N$_{q}$)
indicates the number of k-point used for electron (phonon) momentum integration.}
\label{tab:elph}
\end{table}

\subsection{CaC$_6$\label{cac6}}

Graphite intercalated compounds represent a wide class of compounds. Due to the particular bonding properties of carbon atoms, with strong and stable $sp^2$ bonds and weak Van der Walls $\pi_z$ bonds, graphite allows incorporation of many different atomic species between the hexagonal C sheets.
Alkali metals and alkaline-earth metals are the most common, but even H, Hg, Tl, Bi, As, O, S and halogens are reported\cite{emery}

The possibility to change the number of electrons in the covalent C-C bonds, always stimulated the search of superconductivity in intercalated graphite, but only in  2005 calcium intercalated graphite (CaC$_6$) was found to superconduct at temperatures as high as 11.5 K\cite{weller}.

First-principles theoretical investigations on  CaC$_6$ revealed that the superconducting critical temperature can  be explained considering only the electron-phonon mechanism\cite{CalandraPRL05, SannaPRB07}, even if other pairing mechanisms were proposed (excitonic and plasmonic).
  
The pairing originates from the  electron-phonon interaction between intercalant Fermi Surface and the  C out-of-plane modes and 
Ca in-plane modes\cite{CalandraPRL05}.
These last modes, with very low frequencies ($\simeq$ 15 meV), account for about one-half of the total $\lambda$. 

In this section, we re-examine the phonon and electron-phonon properties of CaC$_6$, by means of the computational framework based on the Wannier interpolation. 

In fact, from a computational point of view, the calculation of the energy position and {\bf q}-dispersion of the low energy Ca modes 
requires very high precision, not obtainable with the state-of-the-art phonon and electron-phonon calculations.

CaC$_6$ is well described by the clean limit regime.
Indeed, the CaC$_6$ zone-center phonon frequency measured by Raman,\cite{RamanCaC6_07,RamanCaC6_09,DeanNA} is 
significantly larger than the DFT adiabatic one but agrees very well 
with the DFT clean-limit-non-adiabatic frequency\cite{Saitta,RamanCaC6_07,RamanCaC6_09,DeanNA}. 
For this reason we will discuss both the adiabatic and the non-adiabatic dispersion.

\subsubsection{Technical details}
\begin{figure}
\includegraphics[width=0.8\columnwidth]{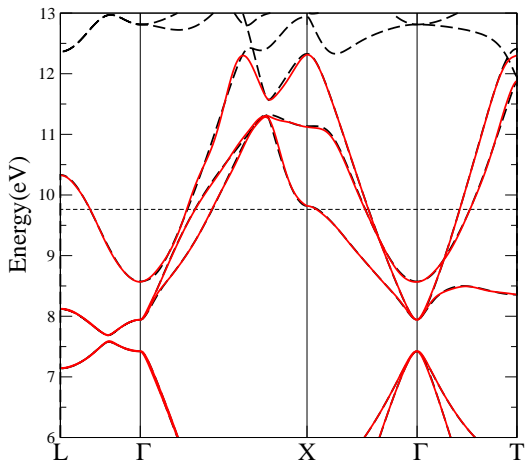}
 \caption{(color online) Wannierized (red-dashed) and first-principles-calculated bands (black-continuous) for CaC$_6$. 
The horizontal dashed line marks the Fermi level.}
\label{fig:bands_CaC6}
 \end{figure}

Wannier-interpolation of the CaC$_6$ band-structure is carried out using a $N_{k}^{w}=6\times 6\times 6$
k-points grid centered at $\Gamma$. We use 7 Wannier functions  starting from an
initial configuration with one s-state on Calcium and one p$_z$ state on each Carbon.
The Wannierized band structure as compared to first-principles calculations 
is shown in Fig. \ref{fig:bands_CaC6}.

We perform one linear response calculation for each point of the irreducible  
$N_{k}^w$ $k-$point grid using $N_{k}(T_{\rm ph})=8\times 8\times 8$ and Hermitian-Gaussian 
smearing $T_{\rm ph}=0.05$ Ryd. We then obtain  $\delta v_{\rm SCF}/\delta \u_{\q s}$
on the full grid applying the symmetry operations of the crystal.
At zone center, we added a Fermi level shift to the $\q=0$ component
of $\delta v_{\rm SCF}/\delta \u_{\q s}$ 
to obtain an analytical 
behavior of at $\q=0$
(see section \ref{sec:efshift} for more details).

The interpolation of phonon frequencies is performed along the 
strategy outlined in sec. \ref{sec:interp_all} and using 
 $T_{\infty}=0.075$ Ryd 
to obtain smoothed ``high temperature''
phonon frequencies. For the calculation of non-adiabatic phonon frequencies we chose
$\omega$ in Eq. \ref{eq:CA_best} as the E$_{2g}$ phonon frequency at 
$\Gamma$ and we also use $N_{k}(T_0)=50^{3}$, T$_{0}=0.01$ Ryd and $\eta=0.01$ Ryd in
Eq. \ref{eq:Pidef}.  

\subsubsection{Adiabatic phonon dispersion \label{sec:CaC6_ph}}

\begin{figure*}[t]
\includegraphics[width=0.7\columnwidth]{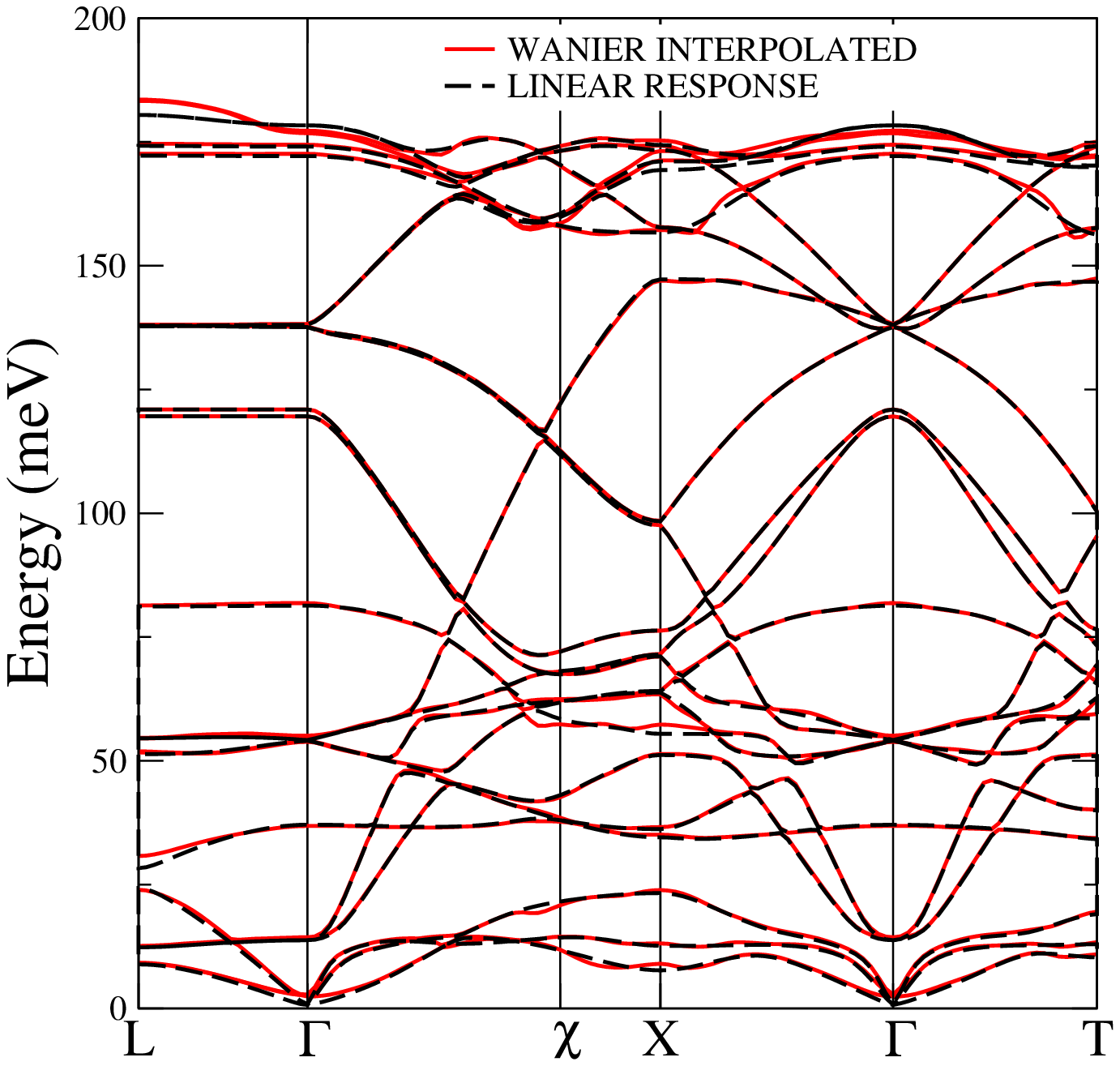}%
\includegraphics[width=0.7\columnwidth]{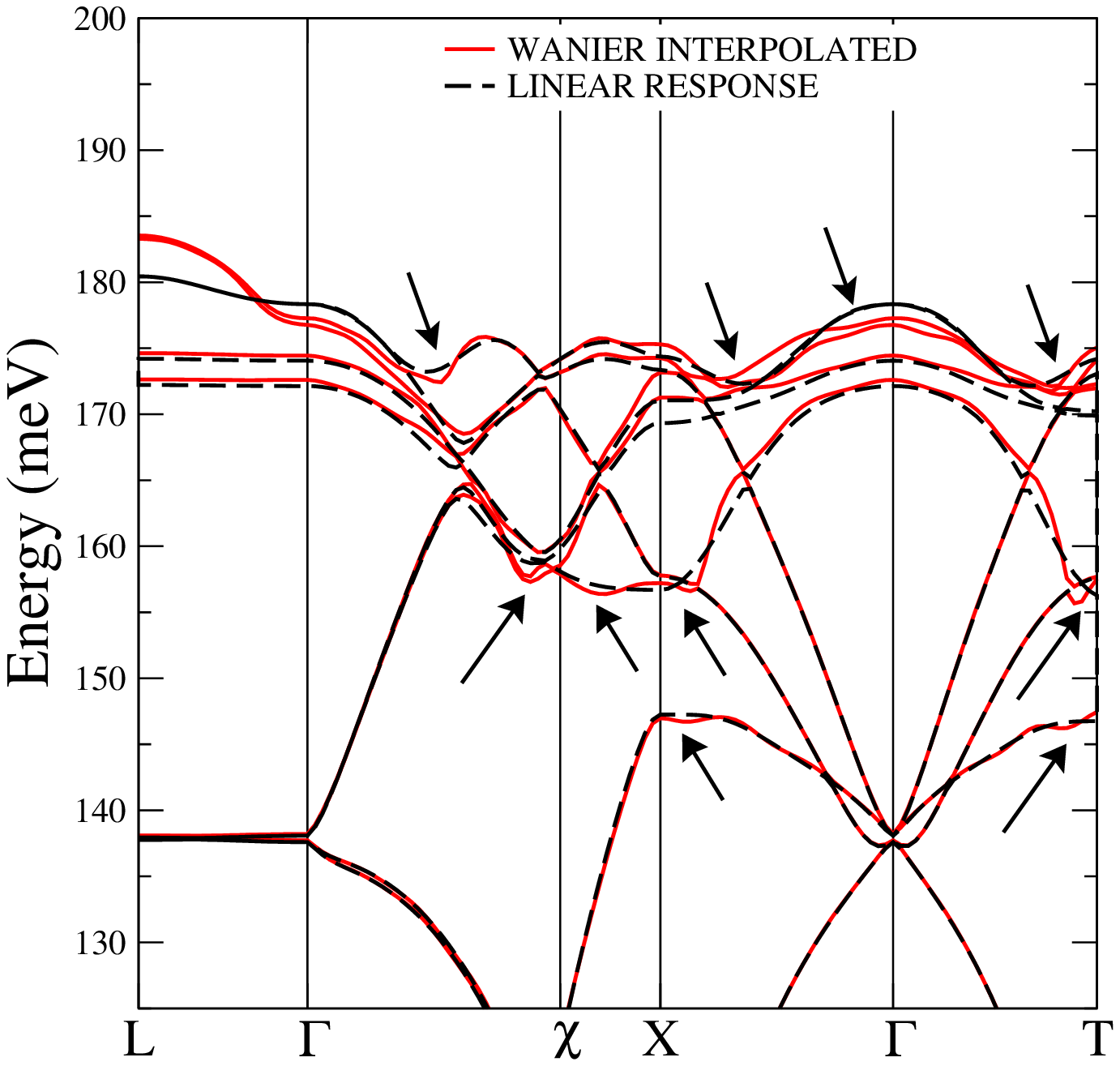}%
\includegraphics[width=0.7\columnwidth]{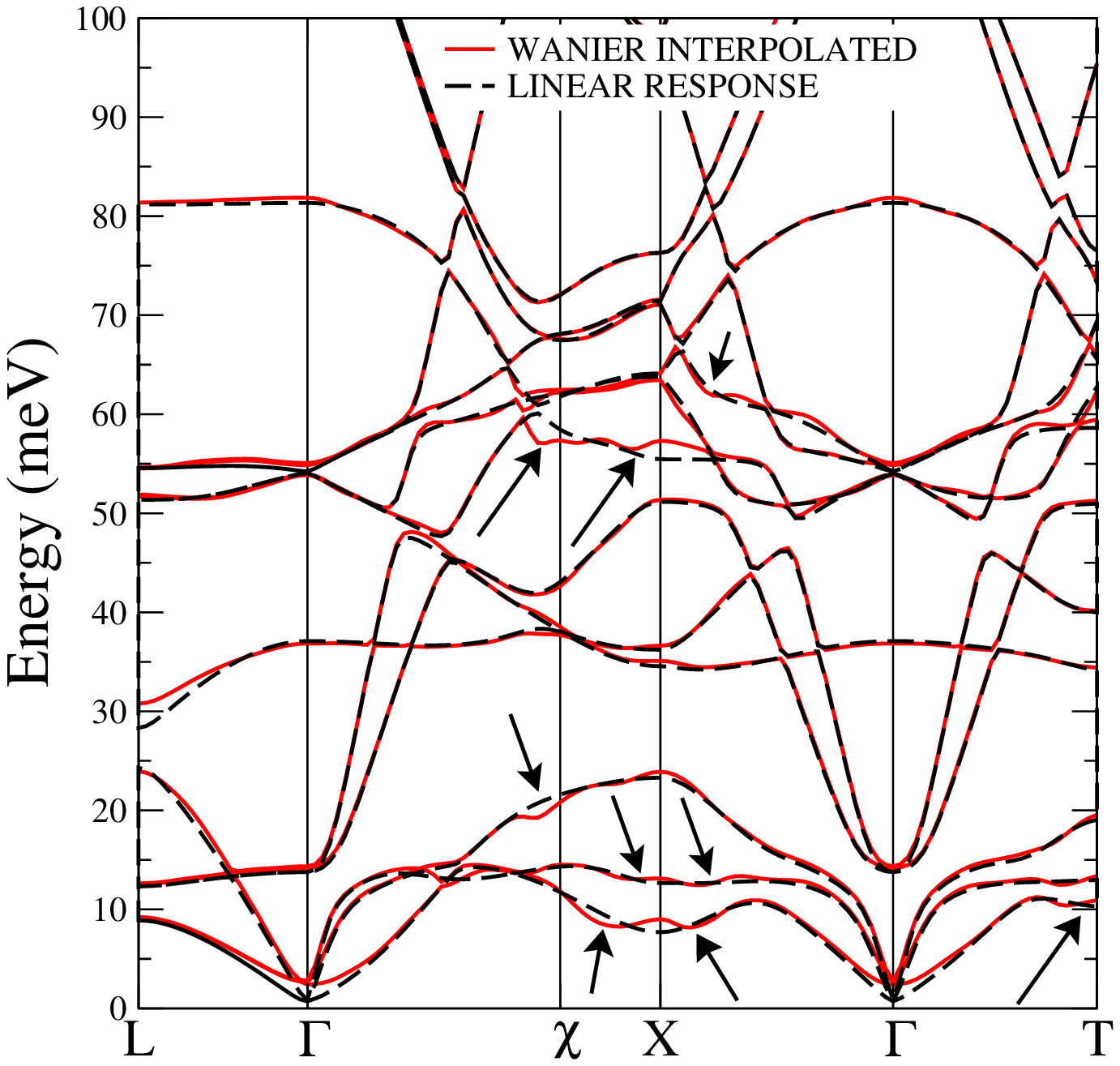}%
 \caption{(color online) Wannier-interpolated (red-continuous) and 
Fourier-interpolated (black-dashed) adiabatic phonon dispersion in CaC$_6$. 
The center and right
panels are zooms to the high and low energy parts of the phonon dispersion.
The Fourier-interpolated phonon branches are obtained from the dynamical matrices
calculated on a $N_{k}^{w}=6^3$ phonon grid, using $N_{k}=8^3$ and Hermite-Gaussian smearing $T_{\rm ph}=0.05$ Ryd. 
The Wannier interpolated bands are
obtained on a $N_{\tilde k}=40^3$  
and Hermitian-Gaussian smearing $T_{\rm ph}=0.01$ Ryd.
The Kohn-anomalies are indicated by arrows.}
\label{fig:AD_dispersion_CaC6}
\end{figure*}

The phonon dispersions of CaC$_6$ as obtained from Fourier-interpolation and 
from Wannier-interpolation are illustrated in Fig. \ref{fig:AD_dispersion_CaC6}.
The Fourier interpolate phonon dispersion is in agreement with previous 
linear response calculations \cite{CalandraPRL05,CalandraPRB06,KimPRB06}.
When compared with the more converged calculations performed using 
Wannier-interpolation several differences are found. 
\begin{figure*}[t]
\includegraphics[width=\columnwidth]{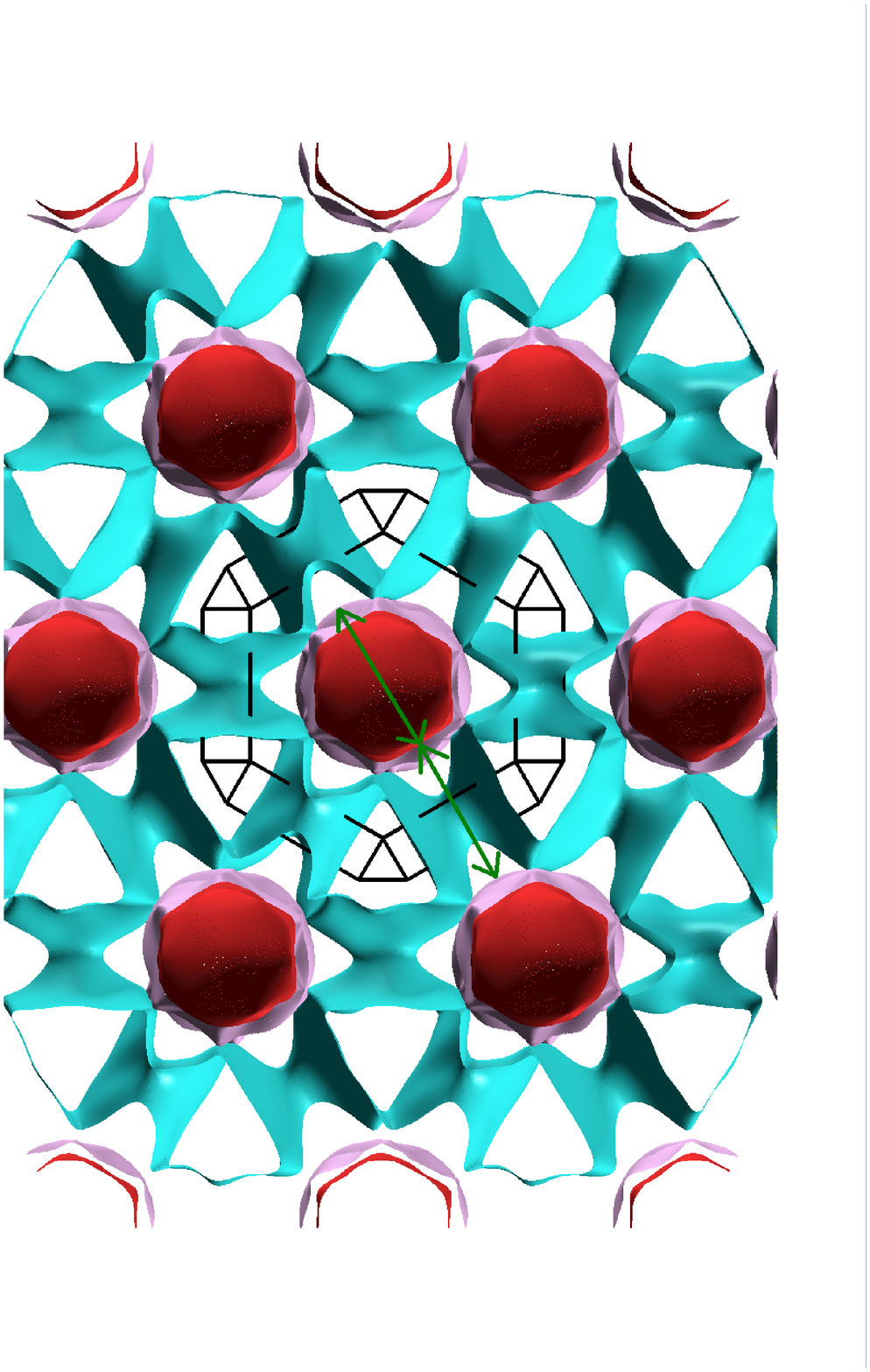}
 \caption{(Color online) Fermi surface of CaC$_6$ calculated using Wannier-functions
on a $100^3$ k-points grid. The green arrows mark the nesting vectors responsible
for the Kohn-anomaly close to X.  }
\label{fig:FS_nesting_CaC6}
\end{figure*}
The largest deviation 
concerns the high-energy modes at $\approx 160-185$ meV, in particular along
$\Gamma L$. These modes involve carbon vibrations parallel to the 
Graphite layers (C$_{xy}$ vibrations). There is a substantial hardening of 
the two highest optical phonon-modes approaching zone-boundary
and a somewhat weaker softening close to zone center. The shape of the 
phonon dispersion of the two highest optical modes along $\Gamma L$ recalls 
the typical behavior of the real-part of the phonon self-energy
due to electron-phonon coupling (see for example for the case of MgB$_2$ Ref. 
\onlinecite{CalandraPRB05}, fig 3a.). The electron-phonon coupling to these 
optical modes comes  
essentially from $\pi^{*}$ states and mostly from intraband-coupling \cite{BoeriJIG} 
(small Fermi surface) so that a quite accurate BZ sampling is necessary to display this
behavior. This explains the strong $k-$point dependence of the result.

Moreover the relevance of having an accurate sampling of the Fermi surface is evident
from the large number of Kohn-anomalies (arrows in Fig. \ref{fig:AD_dispersion_CaC6}
mark the occurrence of a Kohn-anomaly)
present in the Wannier-interpolated branches
and almost always absent in the Fourier interpolated ones.
At high energy (150-180 meV) we count a large number of anomalies,
no one of which is present in Fourier 
interpolated branches. 
These anomalies are mostly located close to the X and $\chi$ points.
At low energy we count 8 more softenings, 6 of which involve points close to
the X-point (although on different branches) and two occur at the $\chi$ point. 
In Fourier interpolated calculations
only a very broad anomaly is visible in the low energy modes, located exactly at the 
X-point, nothing in other energy modes.
The facts that close to the X point
(i) several anomalies at the same vector occur on different modes and that 
(ii) the phonon frequencies
associated to the anomaly are strongly dependent on the electronic temperature 
suggest an origin dependent from the Fermi surface geometry.
Indeed this is the case and the anomaly is illustrated in the nesting vector
in Fig. \ref{fig:FS_nesting_CaC6}.

\subsubsection{Non-adiabatic phonon dispersion \label{sec:CaC6_ph}}

Non-adiabatic effects occur in CaC$_6$ close to zone center
\cite{Saitta, DeanNA}. However it is unknown how strong are these
effects far from zone center,
namely what is the degree of localization of non-adiabatic
effects around the $\Gamma$ point. The question is meaningful as if non-adiabatic
effects extend substantially in the Brillouin zone then they can be relevant
for superconductivity and transport as the average electron-phonon coupling and the
critical temperature could be affected.
\begin{figure}[h]
\includegraphics[width=0.8\columnwidth]{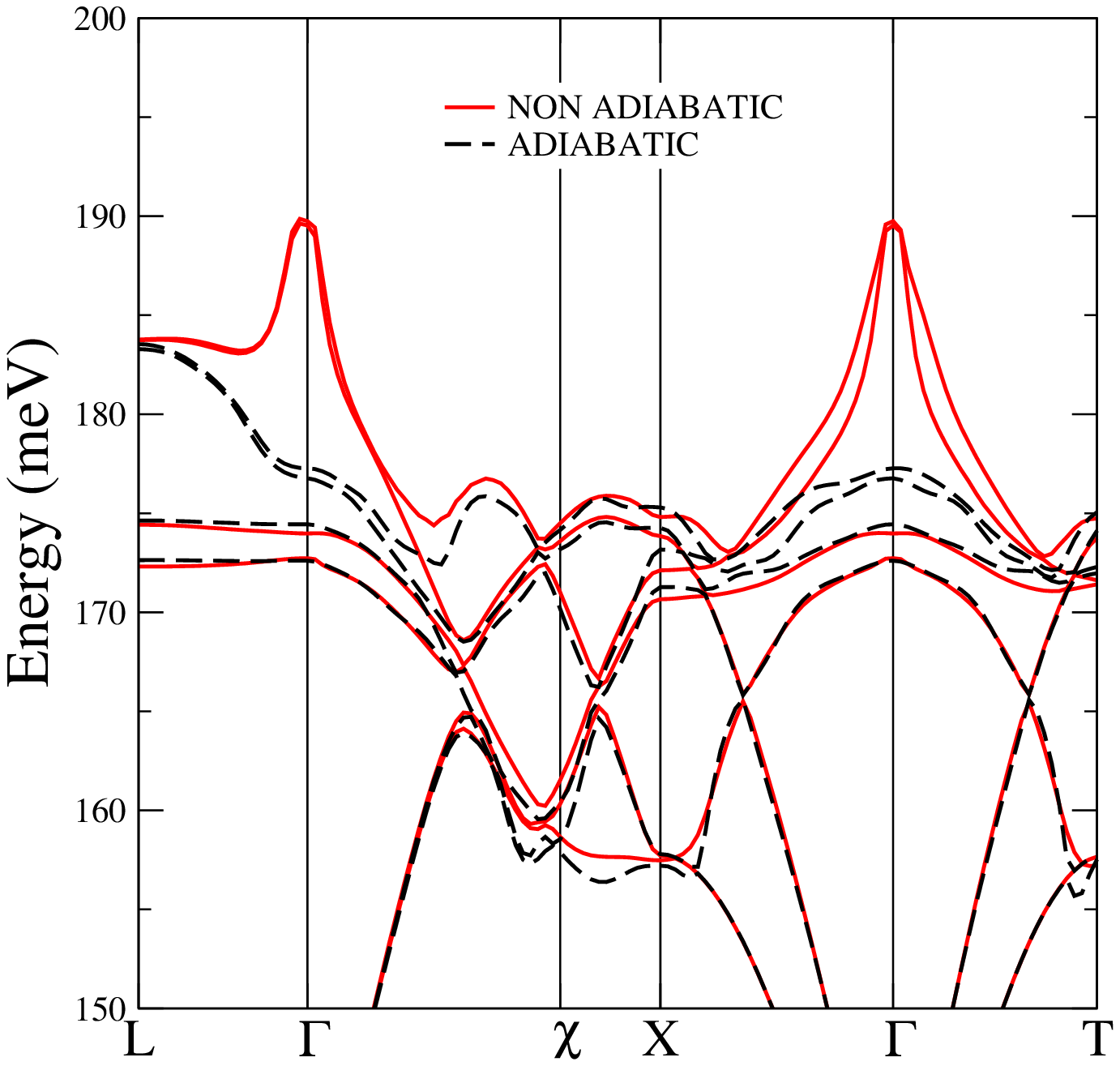}
 \caption{(color online) Wannier-interpolated non-adiabatic (red-continuous) and 
adiabatic (black-dashed) phonon dispersion in CaC$_6$. }
\label{fig:NA_dispersion_CaC6}
\end{figure}

In Fig. \ref{fig:NA_dispersion_CaC6} we plot the Wannier-interpolated 
non-adiabatic phonon dispersion. At $\Gamma$ we obtain $\omega^{NA}=190$ meV
($\approx1532$ cm$^{-1}$), in good agreement with what found in Ref.
\onlinecite{Saitta}. 
Most interestingly the full E$_{2g}$ branch is shifted to higher energies
(particularly along $\Gamma L$, the k$_z$ direction) by the non-adiabatic
effects. Even at half-way to the zone border the correction is 
sizable and it can be easily measured with inelastic X-ray or Neutron scattering
experiments.
In the case of CaC$_6$ as the high-energy modes are weakly-coupled to electrons
the non-adiabatic correction, even if large, probably does not affect the critical
temperature. The situation can be substantially different in other superconductors
so that non-adiabatic effects have to be taken into account to describe
the electron-phonon interaction in layered superconductors. 

\begin{table}
\begin{ruledtabular}
\begin{tabular}{l|c|c|c|c}
{\rm Reference}    & $\lambda$  & $\omega_{\rm log}$(meV) & $N_{k}$ & $N_{q}$ \\ \hline 
Calandra {\it et al.} [\onlinecite{CalandraPRL05}] & $0.83$  &  $24.7$ &  $6\times 6\times 6$ & $4\times 4\times 4$ \\
Kim {\it et al.} [\onlinecite{KimPRB06}] & $0.84$ & $26.28$ & $8\times 8\times 8$ & $4\times 4\times 4$ \\
Sanna {\it et al.} [\onlinecite{SannaPRB07}] & $0.85$ & 28.0 & $6\times 6\times 6 $& $6\times 6\times 6$ \\
This work & $0.829$ & $27.9$  & $59^3$ &$20^3$ \\
\end{tabular}
\end{ruledtabular}
 \caption{Comparison between different electron-phonon coupling calculations 
for CaC$_6$ using different first principles methods.The label N$_{k}$ (N$_{q}$)
indicates the number of k-point used for electron (phonon) momentum integration.}
\label{tab:elphCaC6}
\end{table}

\subsubsection{Electron-phonon coupling \label{sec:CaC6_elph}}

\begin{figure}
\includegraphics[width=0.8\columnwidth]{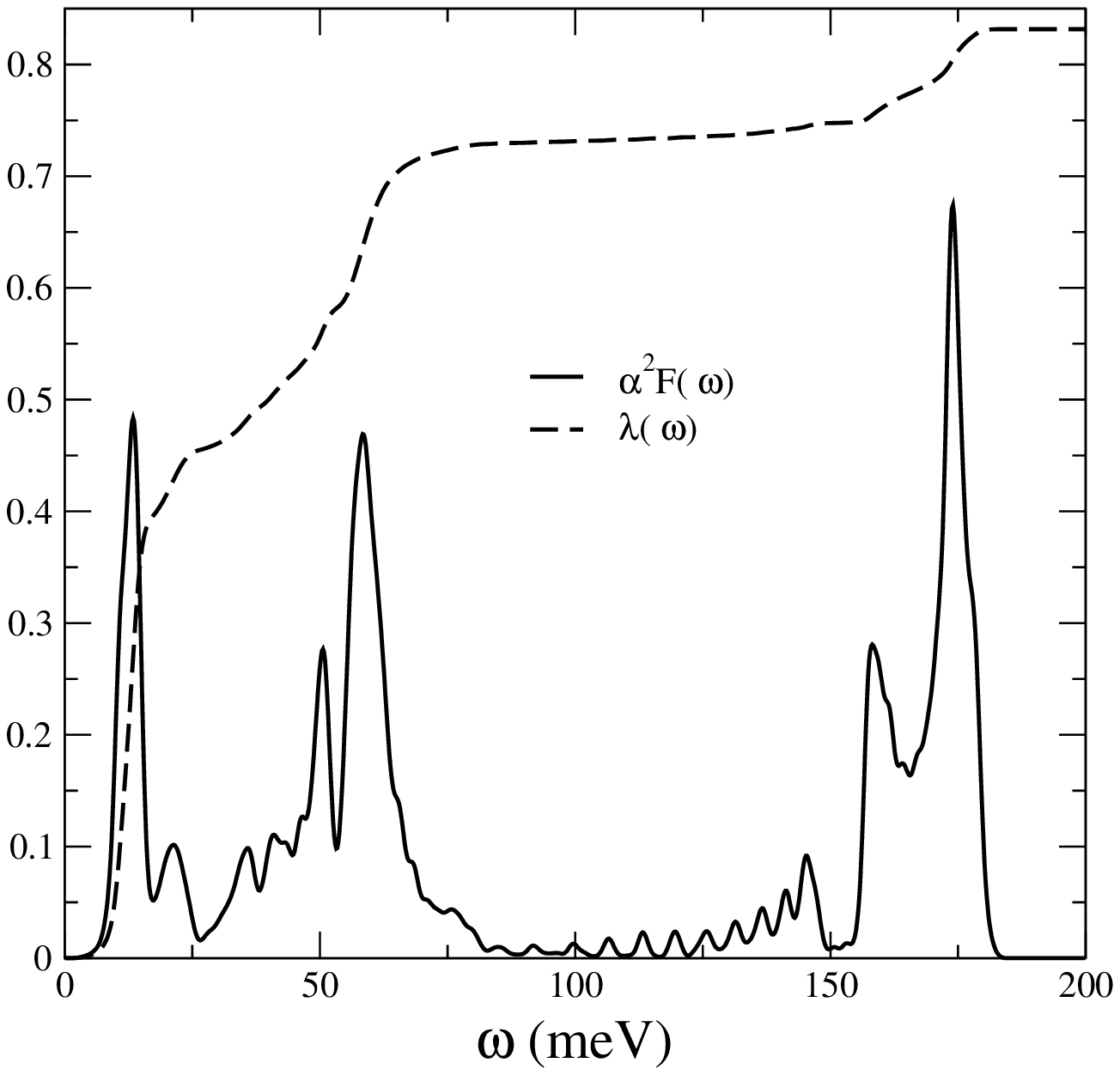}
 \caption{(color online) Wannier-interpolated Eliashberg function and
integrated Eliashberg functions for CaC$_{6}$. Note that the wiggles
in the 80-150 meV region are not due to poor convergence but to the presence
of many well separated dispersive branches in this energy region. }
\label{fig:a2F_CaC6}
\end{figure}

The Eliashberg function calculated with the adiabatic phonon frequencies
and the integrated electron-phonon coupling for CaC$_{6}$ are
illustrated in Fig. \ref{fig:a2F_CaC6}. The situation is very similar to previous
works in what concern the average value of the electron-phonon coupling, as shown in 
table \ref{tab:elphCaC6}. 

Previous calculations performed using standard
linear-response methods report similar values of $\lambda$
and a 13$\%$ variation is present on  $\omega_{\rm log}$ (see table \ref{tab:elphCaC6}).
Our calculation confirms the largest value reported in literature.

\section{Conclusions\label{conclusions}}

In this work we developed a first-principles variational scheme to calculate 
adiabatic and non-adiabatic phonon frequencies in the full Brillouin zone. 
We demonstrated how the method permits the calculation of 
phonon dispersion curves free 
from convergence issues related to Brillouin zone sampling.
Our approach also justify the use of the static screened potential 
in the calculation of the phonon linewidth due to decay in electron-hole pairs.

We apply the method to the calculation of the phonon dispersion and
electron-phonon coupling in MgB$_2$ and CaC$_6$. In both compounds we demonstrate
the occurrence of several Kohn anomalies, absent in previous works, that are 
manifest only after careful electron and phonon momentum integration.
In MgB$_2$, the presence of Kohn anomalies 
on the E$_{2g}$ branches improves the agreement with measured phonon spectra
and affects the 
position of the main peak in the Eliashberg function.
In CaC$_6$ we show that the non-adiabatic
effects on in-plane carbon vibrations are not localized at zone center
but are sizable throughout the full Brillouin zone. 
Thus NA effects can be relevant for the determination of the
superconducting properties of layered superconductors.
In general our work underlines the important of obtaining accurate
(well converged) phonon frequencies to determine the electron-phonon
coupling and superconducting properties in a reliable way.

The large gain in the computational time necessary to obtain phonon 
dispersion provided by our theoretical scheme opens 
new perspectives in large-scale first-principles 
calculations of dynamical properties and electron-phonon interaction.
Furthermore, the variational property of the force constants functional
with respect to the first-order perturbation of the electronic charge 
density is a property that can be generalized to the calculation
of other response functions such as optical properties and the calculation
of Knight shifts as measured in nuclear magnetic resonance\cite{Mayeul07}.

\section{Acknowledgements}
We acknowledge fruitful discussion with J. Yates and support from
ANR PNANO-ACCATTONE. 
GP acknowledge support from CNRS.
Calculations were performed at the IDRIS supercomputing center and at Cineca (Bologna, Italy) through an 
INFM-CNR supercomputing.



%

\begin{thebibliography}{99}
\bibitem{comment} In principle, the force constants matrix depends implicitly on the temperature used 
in the self-consistent calculation of $n_0(\r)$, $\eps_{{\k}, i}$ and $\psi_{\k,i}(\r)$ \ref{eq:DE_DRDR}.
Moreover, the dependence of force constants on this last temperature is weak and thus it is usually taken (as in the present derivation)
equal to $T_{\rm ph}$.
However, it should be set equal to the physical temperature, $T_0$ as a single standard self-consistent calculation 
does not poses  relevant problems in k-point convergence.
\end{thebibliography}
\end{document}